\documentclass[12pt]{iopart}
\usepackage{textcomp}
\usepackage[dvips]{graphicx}
\usepackage[usenames,dvipsnames]{color}

\newcommand {\SPS}    {Sn$_2$P$_2$S$_6$\ }
\newcommand {\SPSe}    {Sn$_2$P$_2$Se$_6$\ }
\newcommand {\SPSSe} {Sn$_2$P$_2$(Se$_x$S$_{1-x}$)$_6$\ }
\newcommand {\PSPS} {(Pb$_y$Sn$_{1-y}$)$_2$P$_2$S$_6$\ }

\begin{document}

\title[Ferroelectricity in \PSPS mixed crystals and random field BEG model]{Ferroelectricity in \PSPS mixed crystals and random field BEG model}
\author{K~Z~Rushchanskii$^1$, R~M~Bilanych$^2$, A~A~Molnar$^2$, R~M~Yevych$^2$, A~A~Kohutych$^2$, S~I~Perechinskii$^2$, V~Samulionis$^3$, J~Banys$^3$ and Yu~M~Vysochanskii$^2$}

\address{$^1$ Peter Gr\"unberg Institut, Quanten-Theorie der Materialien, Forschungszentrum J\"ulich and JARA, D-52425 J\"ulich, Germany}
\address{$^2$ Institute for Physics and Chemistry of Solid State , Uzhgorod National University, 88000 Uzhgorod, Ukraine}
\address{$^3$ Faculty of Physics, Vilnius University, Sauletekio 9/3, LT-10222 Vilnius,
Lithuania}
\ead{ryevych@gmail.com}
\begin{abstract}
For \SPS ferroelectrics the second order phase transitions line is observed until reaching the tricritical point at transition temperature lowering to 250 K by compression. Observed temperature-pressure phase diagram agrees with simulated diagram by MC calculations based on early founded by DFT study~\cite{b1} local potential for \SPS crystals. In addition to the tricritical point, the possibility of disordered and quadrupolar phases occurrence was shown. For mixed crystals with tin by lead substitution, the investigated ultrasound, hypersound and low frequency dielectric properties also reveal appearance of heterophase peculiarities at decreasing of ferroelectric transition temperature below so named "temperature waterline" near 250~K. The tricriticality at similar temperature level also appears in mixed crystals at sulfur by selenium substitution. Such behavior agree with Blume-Emery-Griffiths (BEG) model, that is appropriated for investigated ferroelectric system with three-well local potential for the order parameter (spontaneous polarization) fluctuations.
\end{abstract}

\pacs{62.65.+k, 64.60.-i, 77.84.-s}
\noindent{\it Keywords\/}: ferroelectrics, tricritical point, temperature-pressure phase diagram, MC calculations, Blume-Emery-Griffiths model
\section{Introduction}
\SPS crystals are uniaxial ferroelectrics with three-well local potential for the order parameter fluctuations~\cite{b1}. This potential can be related to the BEG model for analysis of the systems with pseudospin values 0 and $\pm1$~\cite{b2,b3,b4}. Such model predicts possibility of the tricritical point (TCP) presence on state diagrams, where the second order phase transitions (PT) line transforms into the first order transitions line. This point can be observed at transitions temperature lowering by pressure influence or by change of chemical composition. In addition to the first order transition line, at least metastable paraelectric and ferroelectric states are also predicted~\cite{b3}. Moreover, phase diagram could be also complicated, for example, by the quadrupole state occurrences~\cite{b4}.

In the mixed crystals the random bond and random field defects could strongly influence on the phase diagram that is predicted by the BEG model. The coordinates of TCP could be shifted and segment of earlier first order transitions line could be observed as continuous transitions~\cite{b5}. In addition, the dipole glassy or ferroglassy regions could appear on the phase diagram~\cite{b6}.

For \SPS crystals under compression, the second order ferroelectric transition obviously change character to first order at $p\approx0.4$~GPa and $T\approx250$~K as follows from the ultrasound temperature anomalies analysis~\cite{b7}. The neutron diffraction data show~\cite{b8} that under pressure of 0.6~GPa the transition from paraelectric into ferroelectric phase definitely is the first order. The high resolution X-ray diffraction temperature measurements at different pressures also clearly demonstrate the first order character of ferroelectric transition in \SPS crystals under pressure of 1.2~GPa where the transition temperature is lowered till 110~K~\cite{b9}.

At sulfur by selenium substitution in \SPSSe mixed crystals, the TCP on the phase diagram is located at $x\approx0.6$ and $T\approx240$~K~\cite{b10}. Here the TCP is a virtual one because the incommensurate phase appears at $x\geq x_{LP}\approx0.28$ ($x_{LP}$ is the Lishitz point coordinates)~\cite{b11}.

The three-well potential in  \SPS crystals is related to the stereoactivity of Sn$^{2+}$ cations 5s$^2$ electron lone pair~\cite{b1}. At sulfur by selenium substitution, the chemical bonds covalence increases what determines weaker intercell interactions at obviously similar shape of the local three-well potential. Herewith the temperature of ferroelectric transition lowers. At compression the chemical bonds ionicity rises and stereoactivity of Sn$^{2+}$ cations weakens. This fact determines strong transformation of the local three-well potential -- the side wells become more flat what naturally lowers the temperature of ferroelectric transition and prompts reaching the TCP near 250~K.

It is interesting to understand the ferroelectric transition evolution at tin by lead substitution in \PSPS mixed crystals. In a case of PbTiO$_3$ crystal, the stereoactivity of Pb 6s$^2$ electron lone pair together with the covalence of Ti--O bonds determines the strongest ferroelectricity in the family of oxide perovskites. However, in chalcogenide materials the energetic distance between Pb 6s orbitals and S 3p orbitals is a bigger in compare with a distance to energy levels of O 2p orbitals. So, naturally the stereoactivity of lead two charged cations in chalcogenides is expected to be weaker than in oxide case, and they could be also smaller relatively to the tin cations stereoactivity. Hereby for Pb$_2$P$_2$S$_6$ compound, the paraelectric state could be stable until the lowest temperatures. At first structure determination~\cite{b12}, the P2$_1$/c space group was found for Pb$_2$P$_2$S$_6$ crystal, but the acentric Pn structure have been proposed at room temperature later~\cite{b13}. By experimental pyroelectric and piezoelectric testing, it was supported the former conclusion about centrosymmetric structure of Pb$_2$P$_2$S$_6$ crystal at room temperature~\cite{b14}. In addition, the Raman scattering spectroscopy shows some growth of Pb$_2$P$_2$S$_6$ lattice anharmonicity at cooling -- the frequency of the lowest energy optic phonons lowers a little at cooling from 200~K to 77~K~\cite{b14}. This founding could be interpreted as possibility of a virtual or very low temperature ferroelectric phase transition in Pb$_2$P$_2$S$_6$ crystal. Consequently, the Pb$^{2+}$ cations obviously have some "residual" stereoactivity in the Pb$_2$P$_2$S$_6$ lattice.

In the \PSPS mixed crystals with lead concentration increasing, the ferroelectric transition temperature goes down to 4.2~K at $y=0.61$~\cite{b14}. The low temperature ferroelectric ordering or possibly the dipole glassy states were evidenced at $y>0.61$ by dielectric investigations also~\cite{b15}. In whole, the \PSPS mixed crystals present a good possibility for the dipole ordering study in system that could be described within BEG model with random fields. In \SPS family crystals, the metal cations are placed in general positions, the tin by lead substitution destroys the inversion symmetry and induces an electric dipoles.

In this paper, the results of Monte Carlo (MC) simulations of pseudospin distributions at different temperatures and hydrostatic pressures are presented for the case of three-well local potential in \SPS family crystals. The temperature-pressure (T--P) diagram is constructed and compared with available experimental data, especially with an information about the TCP reaching at high pressures. For the \PSPS  mixed crystals the low frequency dielectric, ultrasound and hypersound (by Brillouin scattering) investigations are performed, the TCP possibility and phases coexistence on temperature-composition (T--y) phase diagram are evidenced. The obtained experimental data are compared with predictions of BEG model with random field~\cite{b5}. The TCP is reached at $y>0.2$ and $T\approx240$~K. For composition with $y=0.3$, the phases coexistence was already found in temperature range between $T\approx200$~K and $T\approx207$~K. For $y=0.45$ sample, the clear first order PT is observed at $T_c\approx130$~K in cooling regime and $T_c\approx137$~K at heating. Here the temperature interval of phase coexistence is very wide -- from 117~K to 150~K.

\section{MC simulations}
Origin of lattice instability in \SPS was theoretically studied by some of us in combined \textit{ab initio} and finite-temperature calculations~\cite{b1}. We found strongly anharmonic energy profile along the phase transition path with metastable minimum in the vicinity of paraelectric phase.  We derived effective Hamiltonian in the simple pseudo-spin form, where ferroelectric distortion is described in terms of local mode. This Hamiltonian explicitly accounts lattice anharmonicity in the local-distortion self-energy term, short-range nearest-neighbor inter-site coupling, long-range dipole-dipole interaction, elastic energy, and coupling of the ferroelectric distortion with anisotropic stress. Finite-temperature studies were performed in periodic $12\times12\times12$ supercell by single-flip Metropolis Monte Carlo (MC) simulations.

\begin{figure}[!htbp]
\centering
\includegraphics*[scale=0.28]{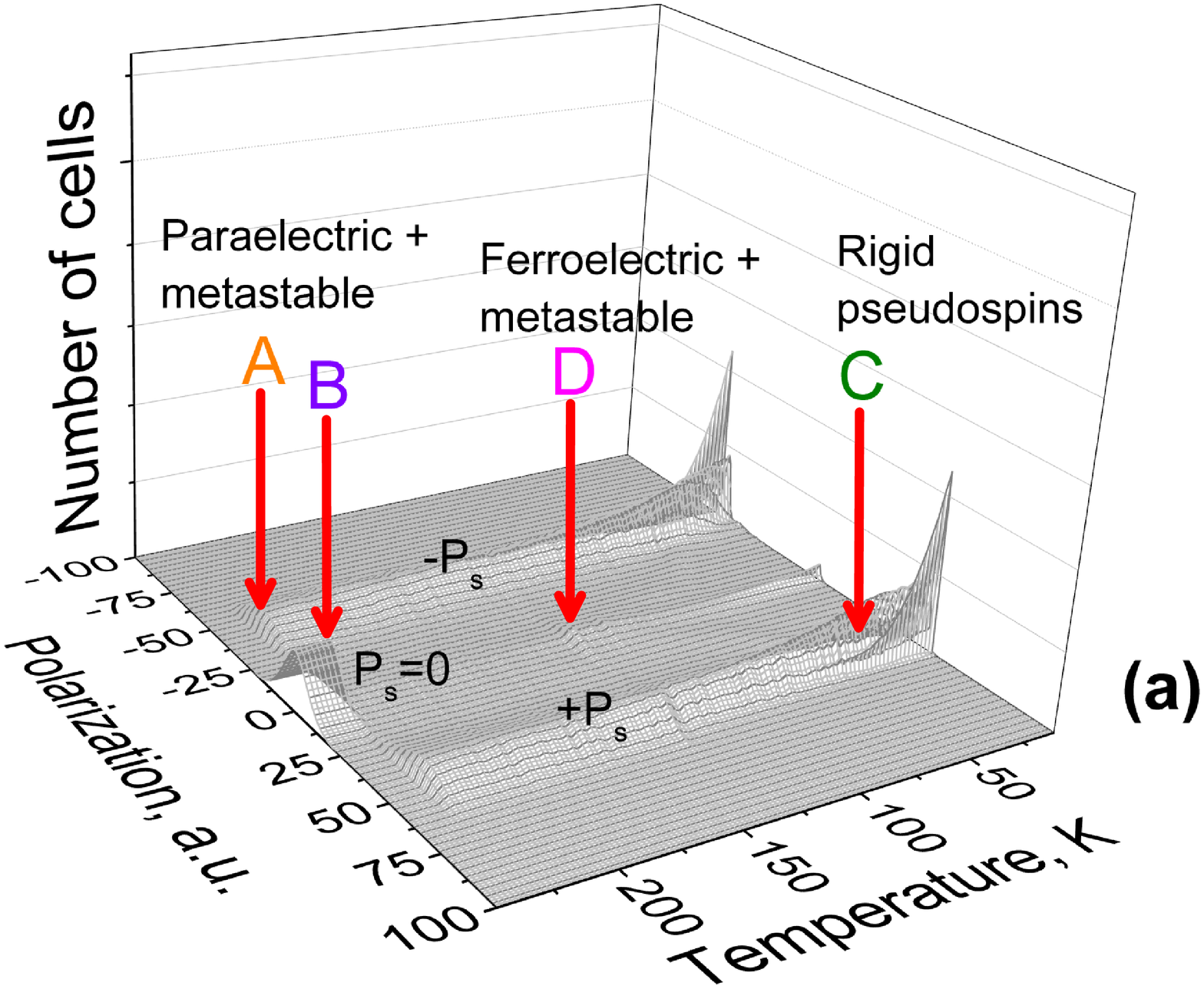}%
\includegraphics*[scale=0.28]{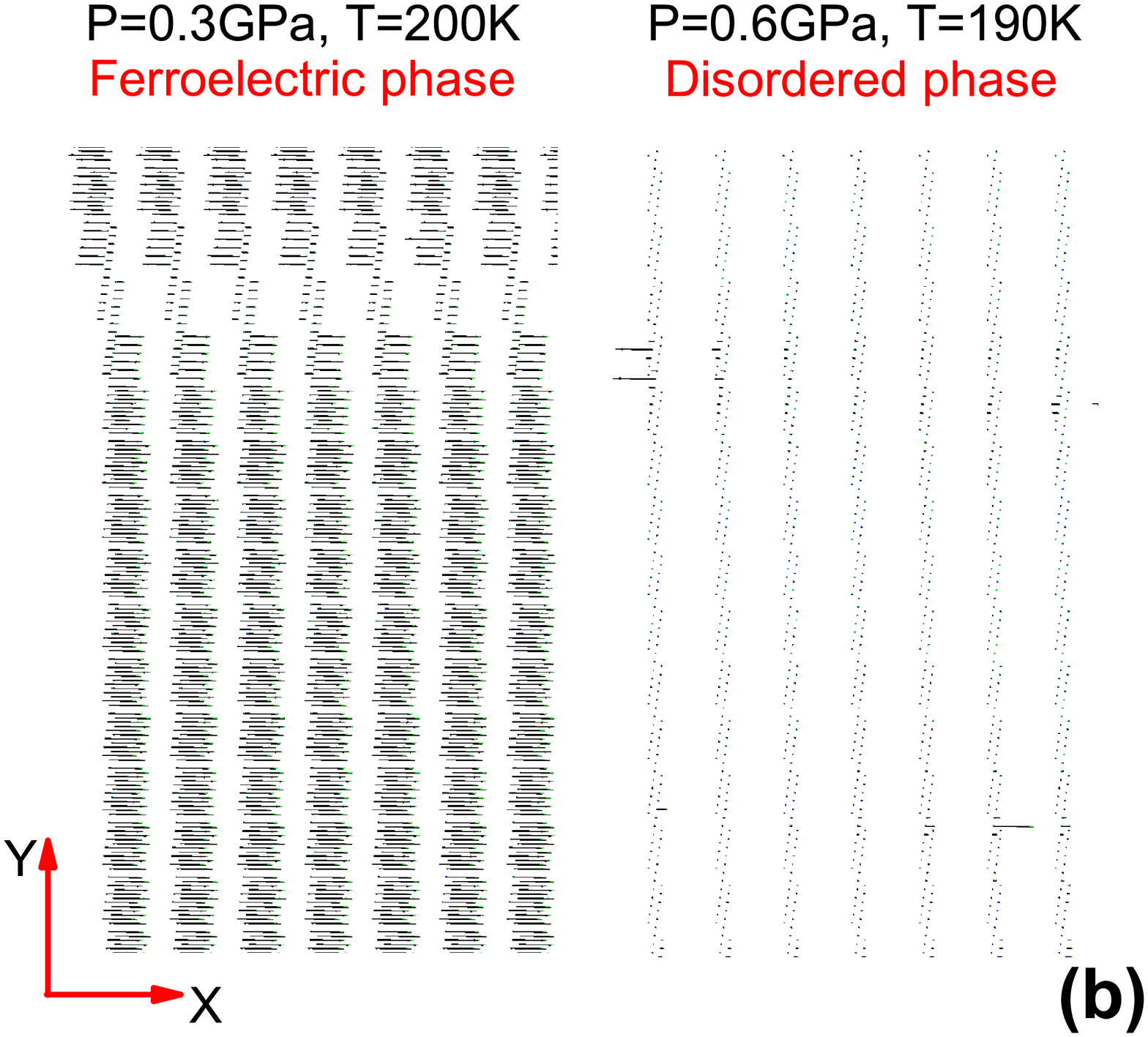}\vspace{1cm}
\includegraphics*[scale=0.28]{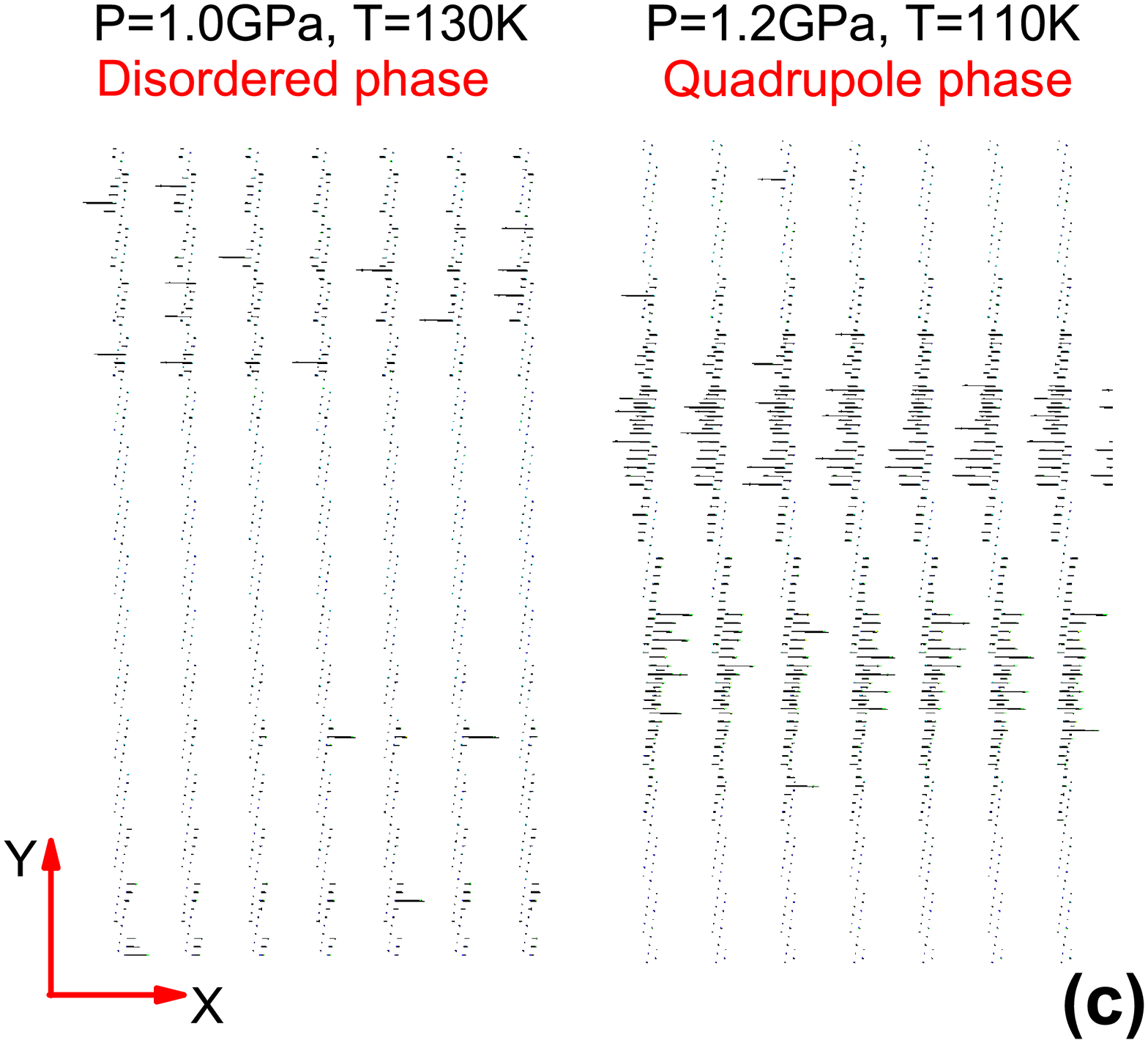}\hfill
\includegraphics*[scale=0.28]{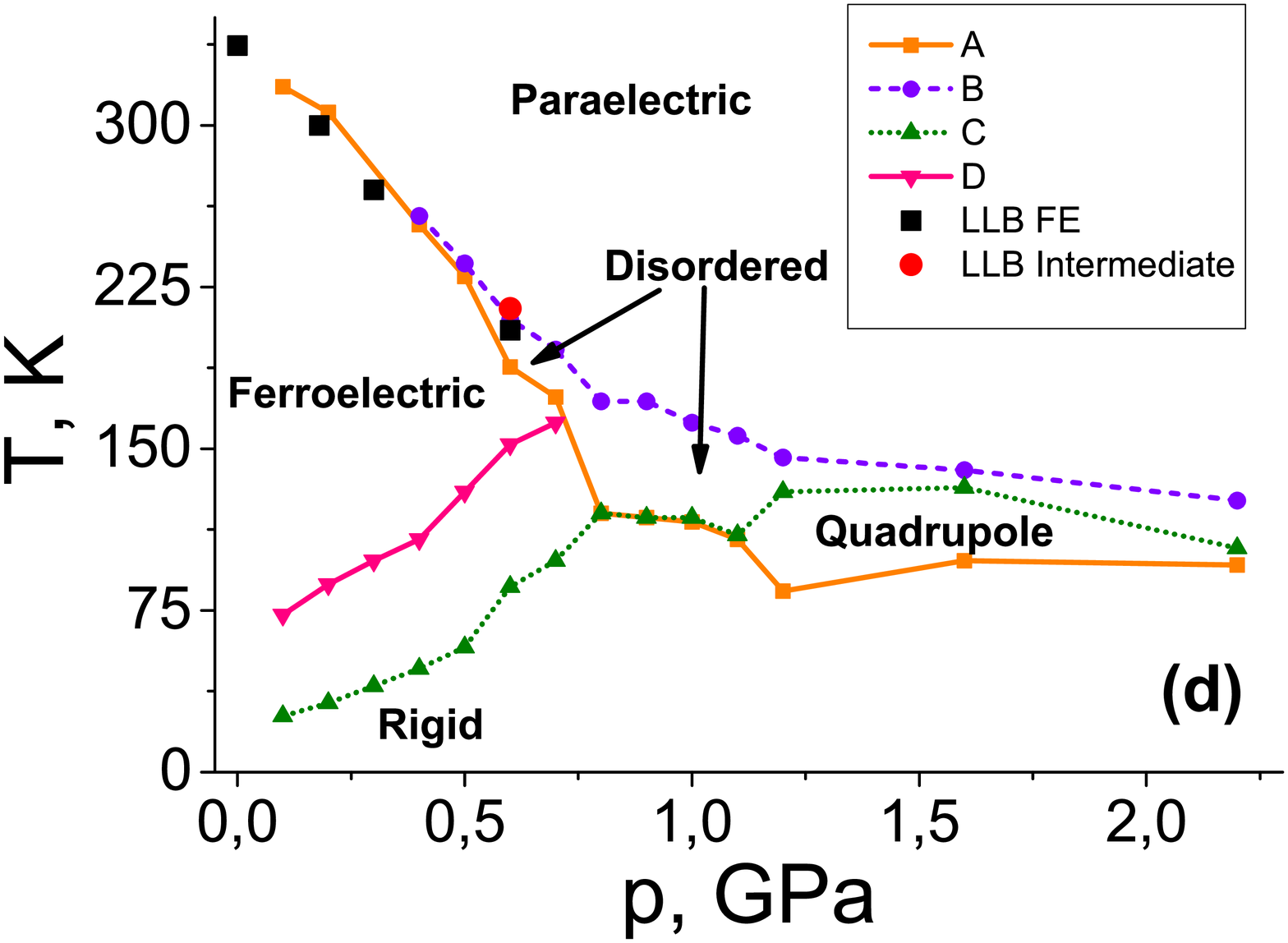}
\caption{Results of MC simulations for the three-well local potential that was determined~\cite{b1} for \SPS ferroelectrics: (a) -- the temperature dependence of pseudospin distribution at $P=0.5$~ GPa; (b), (c) -- pseudospin space distribution in ferroelectric, disordered and quadrupolar phases; (d) --  T--P phase diagram with paraelectric, ferroelectric, rigid, disordered and quadrupole states.} \label{fig1}
\end{figure}

In the present paper we studied temperature-pressure ($T-P$) phase diagram of \SPS with the same Hamiltonian as in the work~\cite{b1}. In order to study possible phenomena, which can appear at the domain boundaries for such anharmonic Hamiltonian, we used periodic $12\times36\times12$ supercell, i.e. elongated along $b$ lattice parameter, which is orthogonal to the domain walls. As in previous study~\cite{b1}, here we used single-flip Metropolis MC method with 10000 MC sweeps per temperature point both for equilibration and production cycles. Initial configuration of pseudospins was constructed as two polar domains with different direction of polarization, which were separated by non-polar area with thickness of one unit cell (see Stable configuration of pseudospins on figure~3 in~\cite{b1}). This configuration was allowed to relax step-by-step in heating regime starting from 10~K and temperature increment of 2~K. The experimental ferroelectric transition temperature was obtained by negative pressure correction $P=-4.3$~GPa.  Results of these simulations are collected on \fref{fig1}.

The T-P diagram (\fref{fig1}(d)) contains phase lines A, B, C and D with the following meaning:

(i) Line A defines the upper limit of the existence of ordered ferroelectric phase. The presence of ferroelectric phase was detected by non-zero number of $\pm P_s$ pseudospins (here $P_s$ is saturated value of spontaneous polarization). The upper boundary of this phase is clearly seen on  temperature diagram of pseudospin distribution on \fref{fig1}(a). Space view of pseudospins in ferroelectic phase is presented on \fref{fig1}(b)(left). Note, that for visibility we focus mainly on domain boundary region, i.e. not complete simulation cell is presented.

(ii) Line B defines bottom limit of the paraelectric phase with zero average values of all pseudospin. Note, that at low pressures the Lines A and B coincide, while above $\sim$0.4~GPa these lines define borders of disordered phase, which has random inclusions of polar cells in the non-polar matrix (see \fref{fig1}(b)(right) as well as \fref{fig1}(c)(left)).

(iii) We found, that at low temperatures some parts of pseudospins have magnitude very close to $\pm~P_s$. When temperature is increased, the number of such pseudospins drops to zero very quickly, whereas the smearing of their magnitude is not influenced. We call this phase as "Rigid pseudospins" one. Line C defines its upper boundary.

(iv) Line D defines the upper limit of metastable domain boundary, which was found and discussed in~\cite{b1}.

In general, obtained theoretical $T-P$ phase diagram of \SPS is unusually rich and complex. At low pressures the direct transition from ferroelectric phase to paraelectric one can be observed. Ferroelectric phase reveals some complex configurations on domain boundaries at low temperatures.

When pressure increases, the ferroelectric phase transition temperature decreases with the slope, which fits well available experimental results \cite{b8,b9} (see label LLB FE and LLB Intermediate on \fref{fig1}(d)).
At pressures higher that 0.4~GPa ferroelectric phase transition line splits and narrow area with chaotic phases appears between purely ferroelectric and paraelectric configurations. Further increase of pressure leads to stabilization of quadrupole phases, as depicted on \fref{fig1}(c)(left). This behavior resembles the theoretical phase diagram of BEG model with negative biquadratic interaction~\cite{b4}. In our theoretical phase diagram the disordered (or chaotic) state clearly appears at pressures $P\geq0.6$~GPa as coexistence of polar and non-polar phases above the first-order ferroelectric phase transition line. Similar coexistence were recently observed as diffuse peaks in neutron and x-ray scattering experiments~\cite{b8,b9}.

As it was obtained in~\cite{b1}, ferroelectric instability in \SPS is a result of non-linear coupling of soft polar mode with low-energy fully symmetrical optical one, which leads to three-well potential. From other side, the system with a three-well potential can be described with two order parameters (which are related to dipole and quadrupole moments). Consequently, this can results in a rich diagram with stable, unstable and metastable states~\cite{b4,b5}. In a real material, the phase diagram could be strongly modified by the lattice imperfection, such as random field defects~\cite{b5}. Our theoretical $T-P$ diagram generally agree with results of recent high-pressure neutron and x-ray diffuse scattering experiments~\cite{b8,b9}. However, further thorough diffraction experiments are still required to evaluate the theoretical $T-P$ phase diagram,  which predicts variety of pressure induced metastable states in \SPS ferroelectrics.

\section{Experimental data}
Dielectric susceptibility was investigated by Goodwill LCR-815 at frequency $10^4$~Hz with temperature variation velocity of 0.1~K/min. The investigated samples were prepared as plates with $5\times5\times3$~mm$^3$ dimensions with silver paste electrodes on largest (001) face that was nearly normal to the spontaneous polarization direction.

The Brillouin scattering spectra were studied using a He--Ne laser and a pressure-scanned three pass Fabry-Perot interferometer with sharpness of 35 and free spectral range of 2.51~cm$^{-1}$. The scattered light in $Y(XX)\overline{Y}$ geometry was collected from the volume of investigated samples. The samples were placed in a UTREX cryostat in which the temperature was stabilized with an accuracy of about 0.3~K~\cite{b18}.

The measurements of the longitudinal ultrasonic velocity and attenuation were performed using a computer controlled pulse-echo equipment~\cite{b19}. The precision of relative velocity measurements was better than 10$^{-4}$. The temperature stabilization was better than 0.02~K. The sample was carefully polished to have precisely parallel faces normal to the $Y$ axis. Silicone oil was used as an acoustic bonds for longitudinal ultrasonic waves. The measurements were carried out at 10~MHz frequency using piezoelectric LiNbO$_3$ transducers. The vapor transport technology was used for investigated \PSPS single crystals growing. For the prepared samples the Cartesian axes $X$ and $Y$ coincide with [100] and [010] crystallographic directions.

\begin{figure}[!htbp]
\centering
\includegraphics*[scale=0.23]{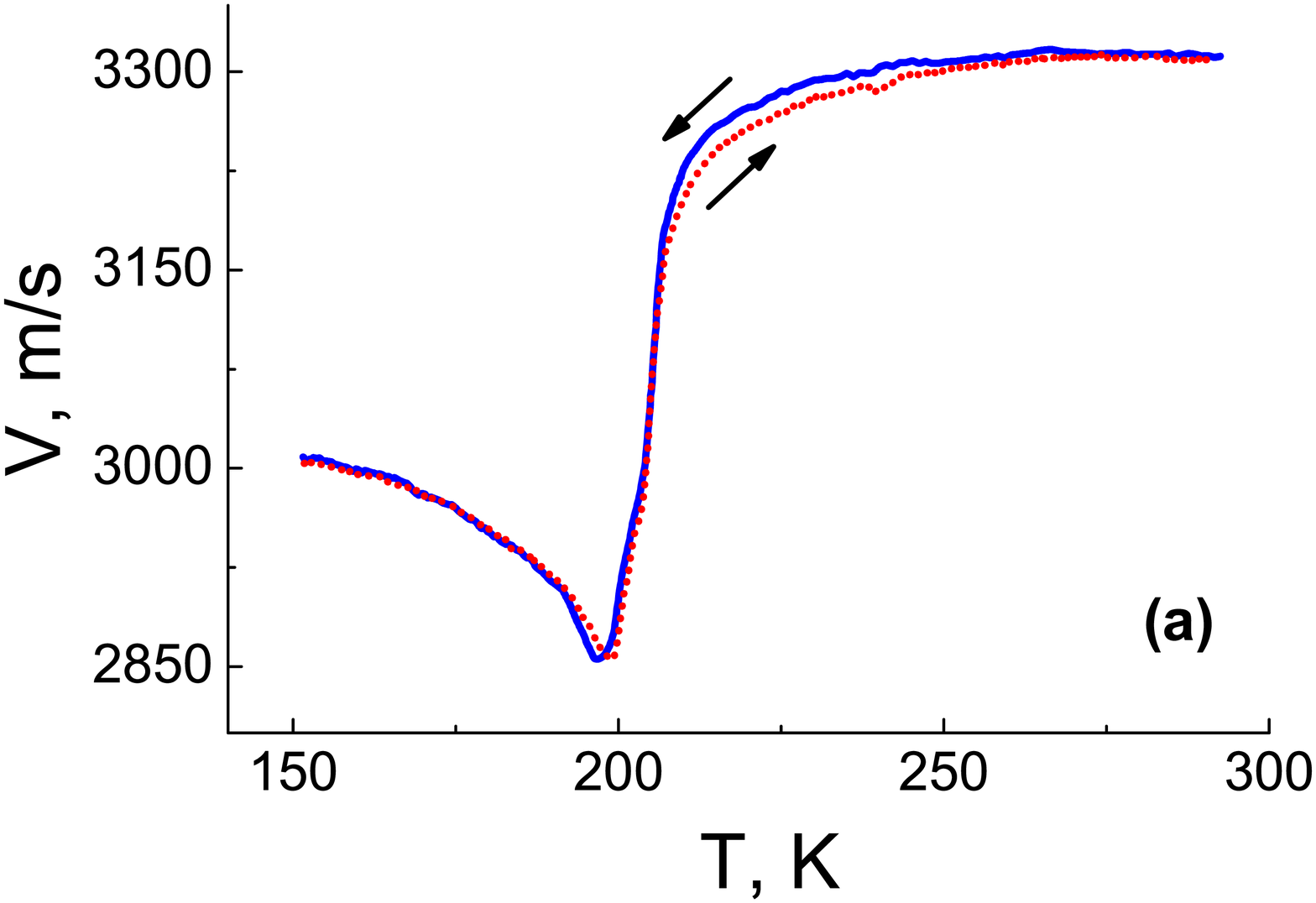}%
\includegraphics*[scale=0.23]{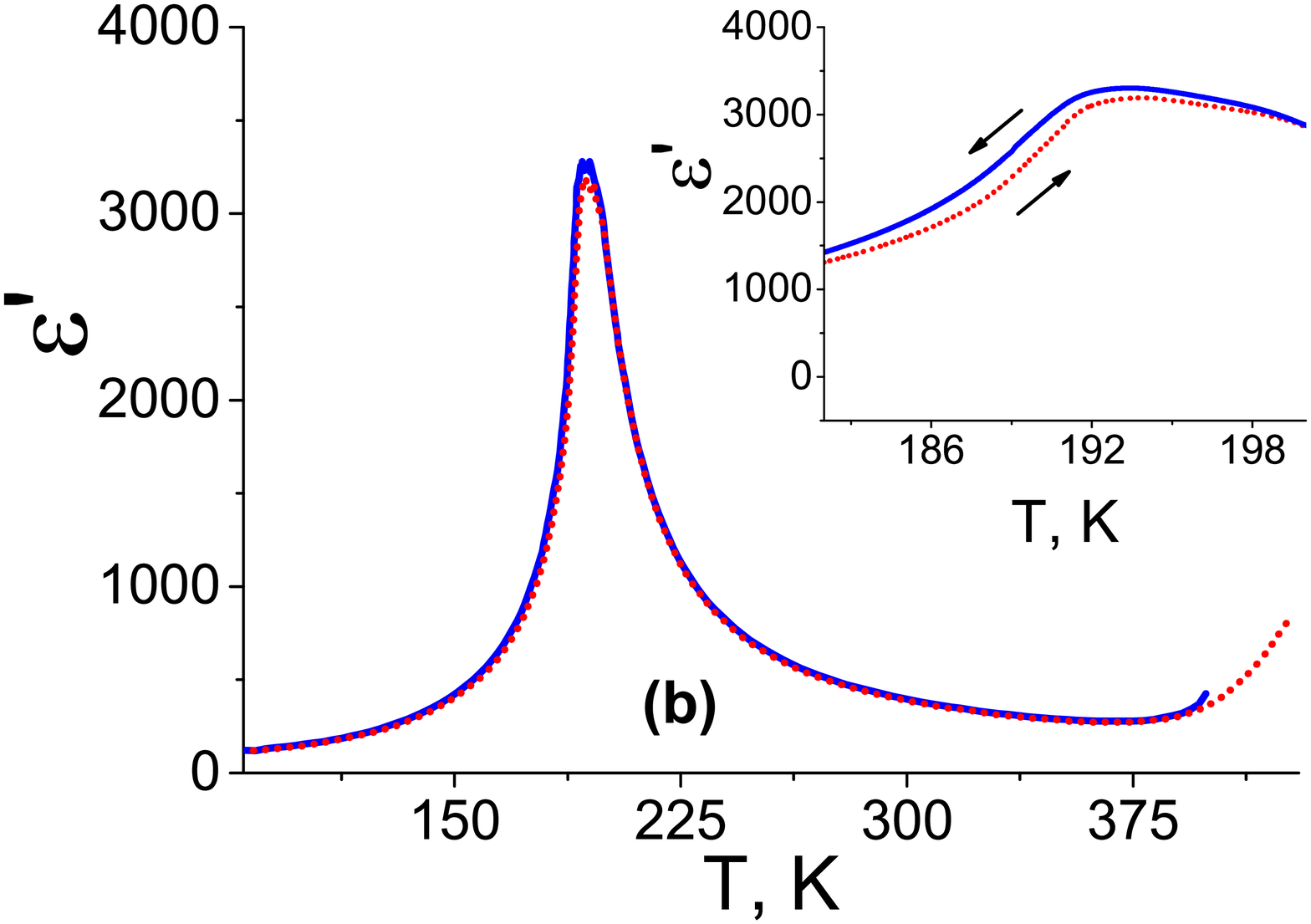}
\caption{Temperature dependencies of ultrasound velocity (a) and dielectric susceptibility (b) at cooling (blue solid lines) and heating (red doted lines) for (Pb$_{0.3}$Sn$_{0.7}$)$_2$P$_2$S$_6$ crystals.} \label{fig3}
\end{figure}

The dielectric susceptibility, ultrasound and hypersound velocity temperature anomalies were used for determination of the ferroelectric phase transition temperature concentration dependence for \PSPS mixed crystals. At rise of lead concentration till $y=0.2$ the temperature of second order phase transition $T_0$ lowers to about 248~K. At $y=0.3$, two anomalies are already observed (at temperatures $T_1\approx200$~K and $T_2\approx207$~K) on ultrasound velocity (\fref{fig3}(a)) and dielectric susceptibility (\fref{fig3}(b)) temperature dependencies.

\begin{figure}[!htbp]
\centering
\includegraphics*[scale=0.23]{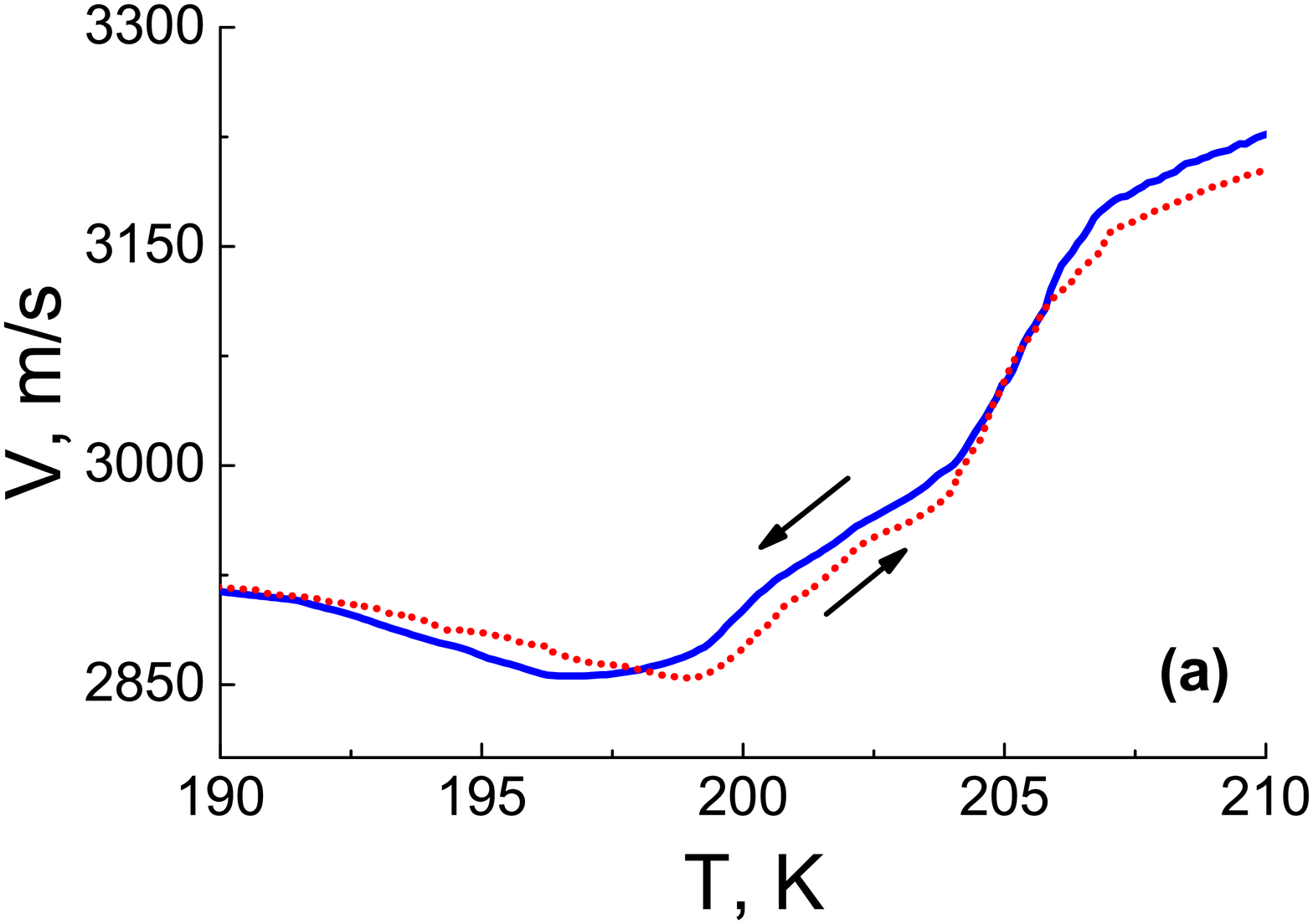}%
\includegraphics*[scale=0.23]{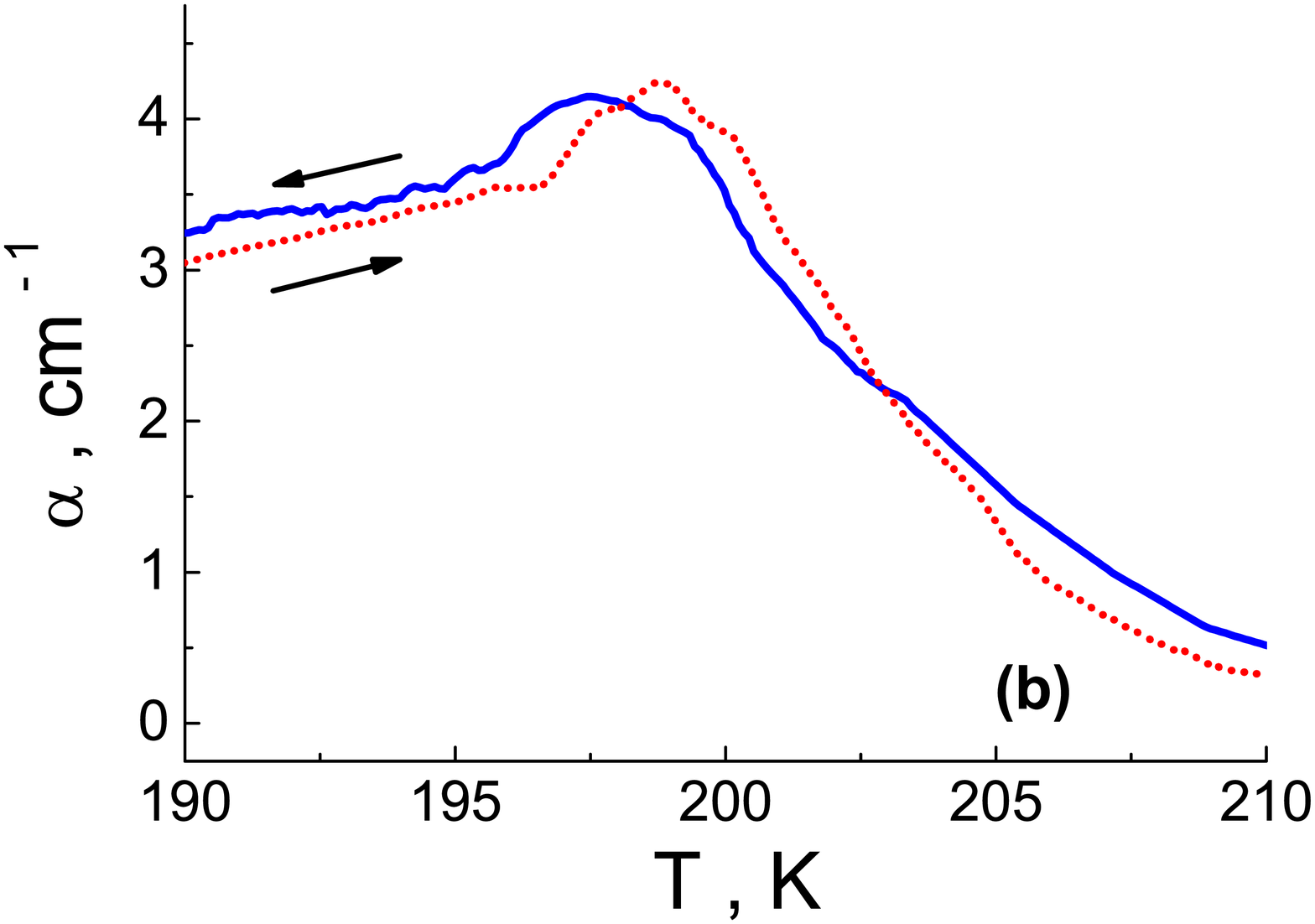}
\caption{Temperature dependencies of ultrasound velocity (a) and attenuation (b) at cooling (blue solid lines) and heating (red doted lines) in the phase transition region for (Pb$_{0.3}$Sn$_{0.7}$)$_2$P$_2$S$_6$ crystals.} \label{fig4}
\end{figure}

For samples with $y=0.1$ and 0.2, the ultrasound attenuation temperature dependence is similar to expected within the Landau-Khalatnikov model for second order phase transitions -- it is seen as asymmetric peak with maximal values near 6~cm$^{-1}$, and with temperature width (at 3~cm$^{-1}$ level) near 2~K and 4~K, sequently (\fref{fig11}). For increased content of lead, in mixed crystal with composition $y=0.3$, the ultrasound anomalies are qualitatively changed -- here addition inflection appears on the step of sound velocity (\fref{fig4}(a)) and attenuation temperature maximum is smeared or delayed into both paraelectric and ferroelectric phases (\fref{fig4}(b)). It could be supposed that first order transition occurs for $y=0.3$ composition already and phase coexistence induce complication of sound velocity anomaly and smearing of attenuation maximum.

\begin{figure}[!htbp]
\centering
\includegraphics*[scale=0.23]{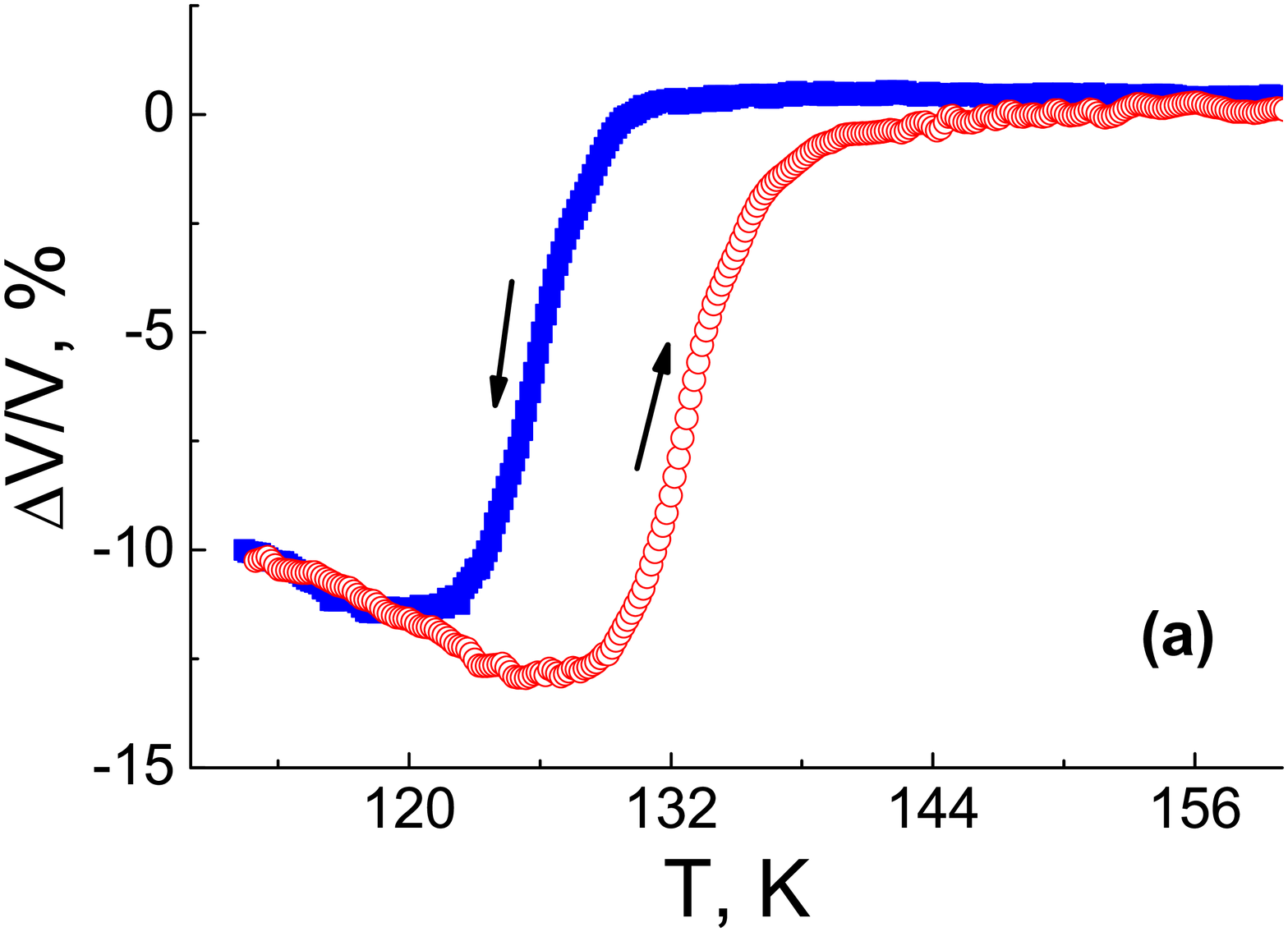}%
\includegraphics*[scale=0.23]{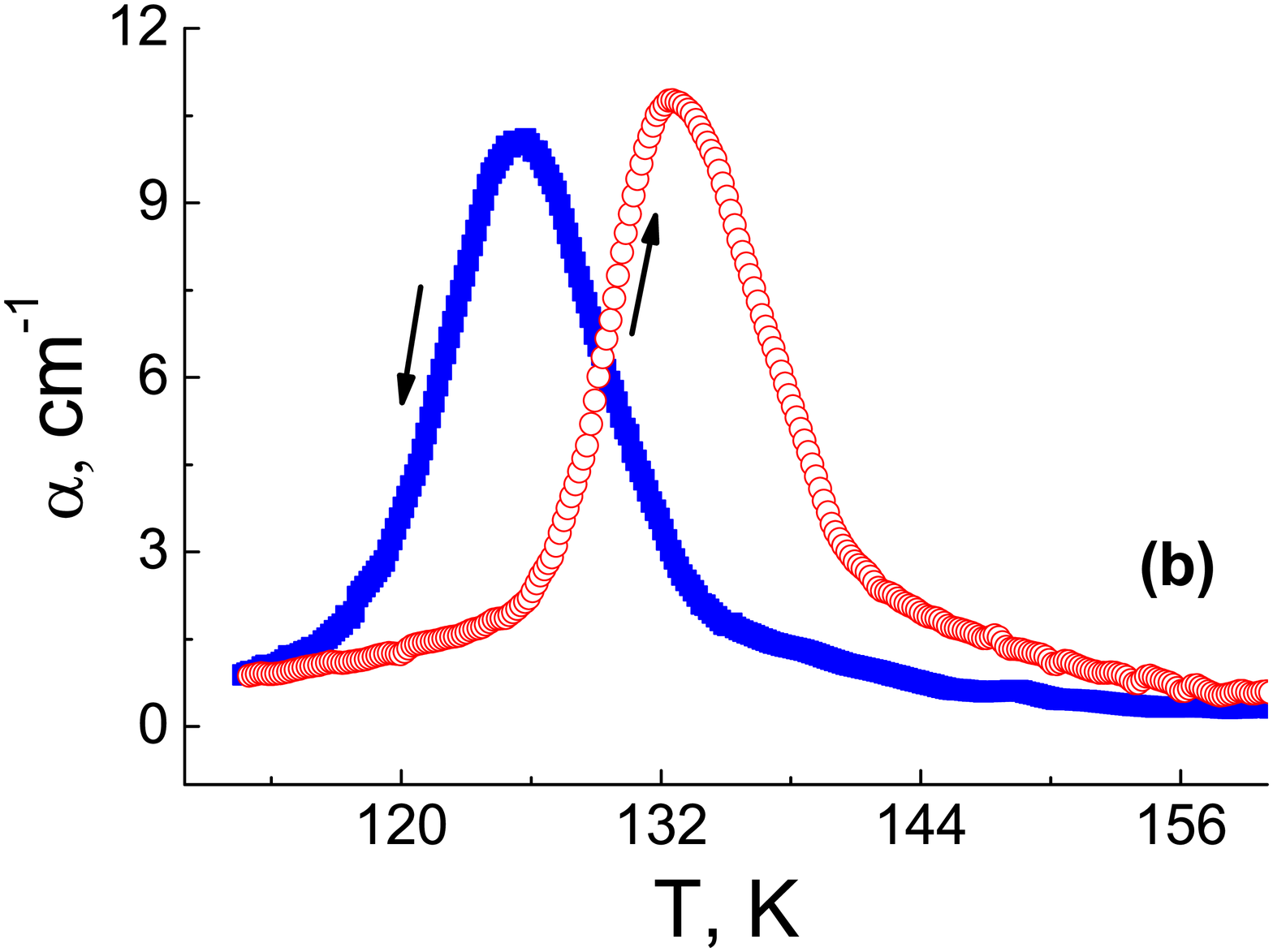}%
\caption{Temperature dependencies of ultrasound velocity (a) and attenuation (b) at cooling (blue solid symbols) and heating (red open symbols) in the phase transition region for (Pb$_{0.45}$Sn$_{0.55}$)$_2$P$_2$S$_6$ crystals.} \label{fig5}
\end{figure}

\begin{figure}[!htbp]
\centering
\includegraphics*[scale=0.28]{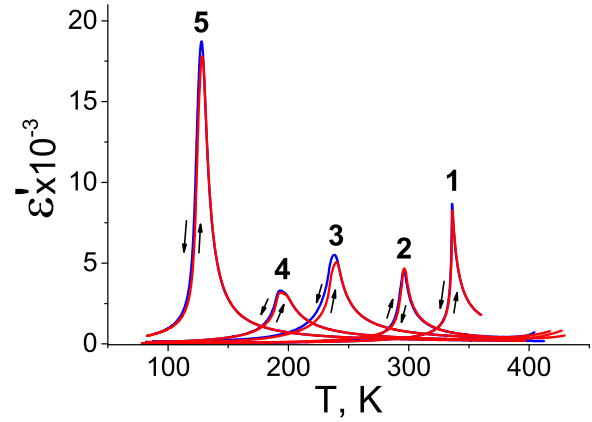}
\caption{Temperature dependence of dielectric susceptibility real part on 10$^4$~Hz at cooling (blue lines) and heating (red lines) for \PSPS \, mixed crystals with $y=0$ -- 1; 0.1 -- 2; 0.2 -- 3; 0.3 -- 4; 0.45 -- 5.} \label{fig6}
\end{figure}

For the $y=0.45$ mixed crystal, the temperature hysteresis of dielectric susceptibility and sound velocity temperature dependencies is clearly observed (\fref{fig5}, \ref{fig6}) what is related to first order character of ferroelectric phase transition near 132~K. For ultrasound velocity and attenuation anomalies, the temperature hysteresis equals about 10~K. In heating mode, the sound velocity and attenuation temperature dependencies have anomalous delay into paraelectric phase within interval above 10~K. Temperature maximum of attenuation attends 10~cm$^{-1}$ at cooling and 11~cm$^{-1}$ at heating. In both regimes, this maximum has temperature width about 30~K.

\begin{figure}[!htbp]
\centering
\includegraphics*[scale=0.28]{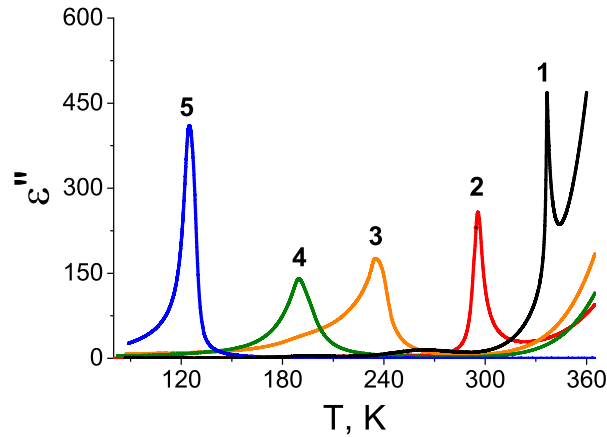}%
\caption{Temperature dependence of dielectric susceptibility imaginary part at cooling on 10$^4$~Hz for \PSPS \, mixed crystals with $y=0$ -- 1; 0.1 -- 2; 0.2 -- 3; 0.3 -- 4; 0.45 -- 5.} \label{fig7}
\end{figure}

The second order ferroelectric transition obviously changes its character to first order at $y>0.2$ what could be seen on transformation of the low frequency dielectric susceptibility temperature anomalies (\fref{fig3}, \ref{fig6}, \ref{fig7}). For sample with $y=0.3$, the anomalies of real and imaginary parts of dielectric susceptibility become wider and have clear temperature hysteresis. For these compositions, temperature dependence of dielectric losses is delayed into ferroelectric phase. For $y=0.45$, the dielectric losses have the biggest value near first order phase transition and their temperature dependence has a shape of almost symmetric maximum (\fref{fig7}).
\begin{figure}[!htbp]
\centering
\includegraphics*[scale=0.23]{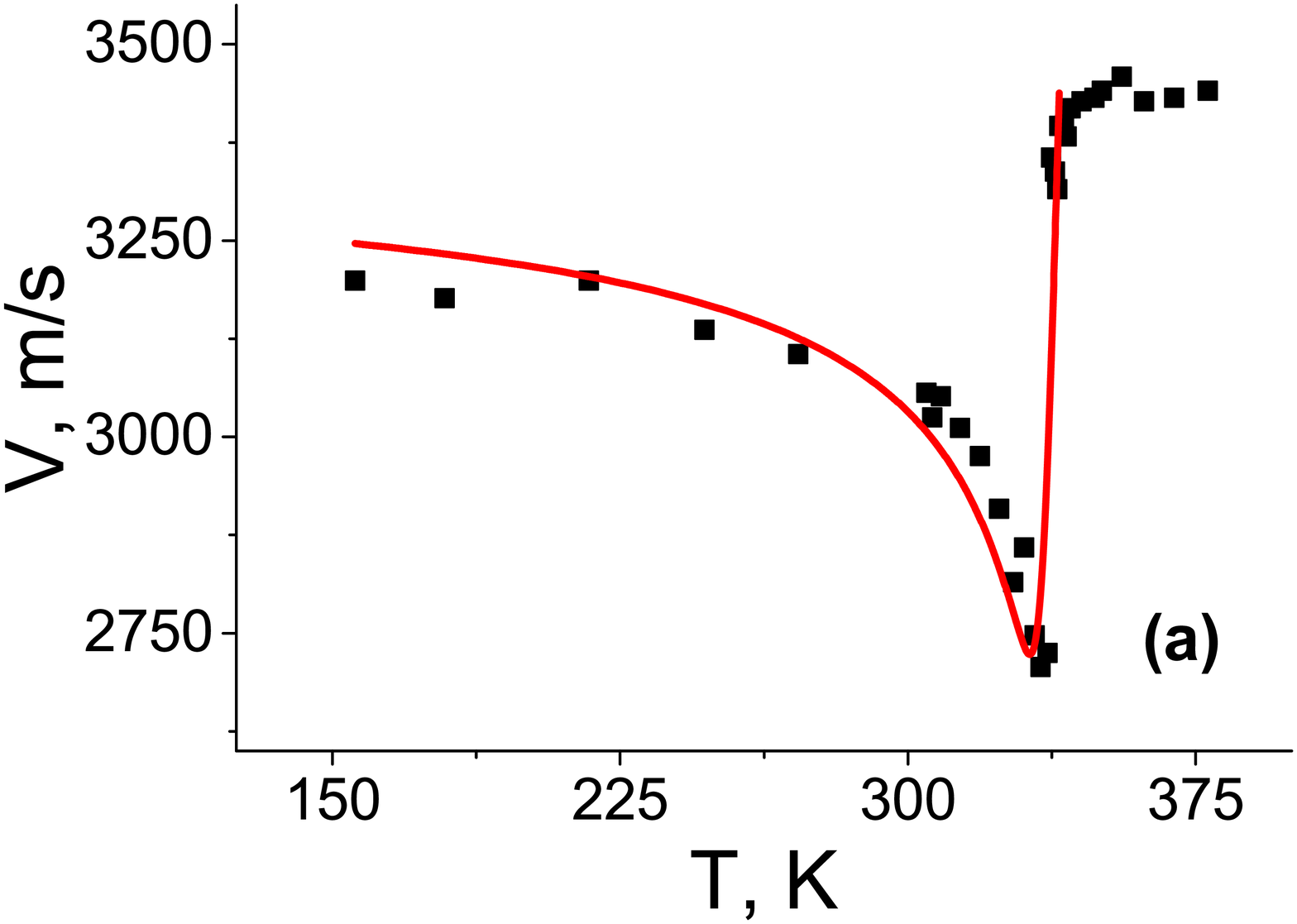}%
\includegraphics*[scale=0.23]{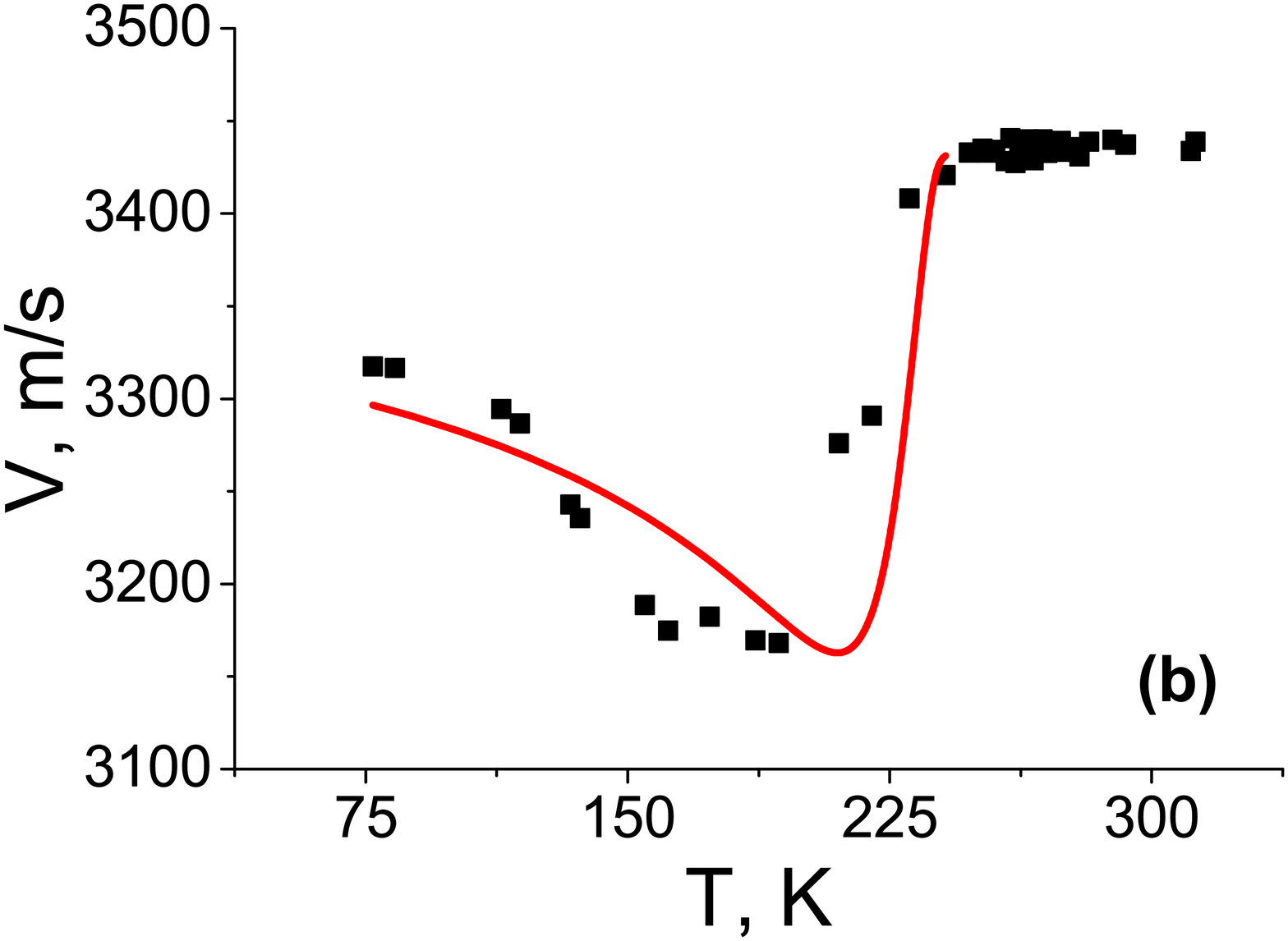}
\includegraphics*[scale=0.23]{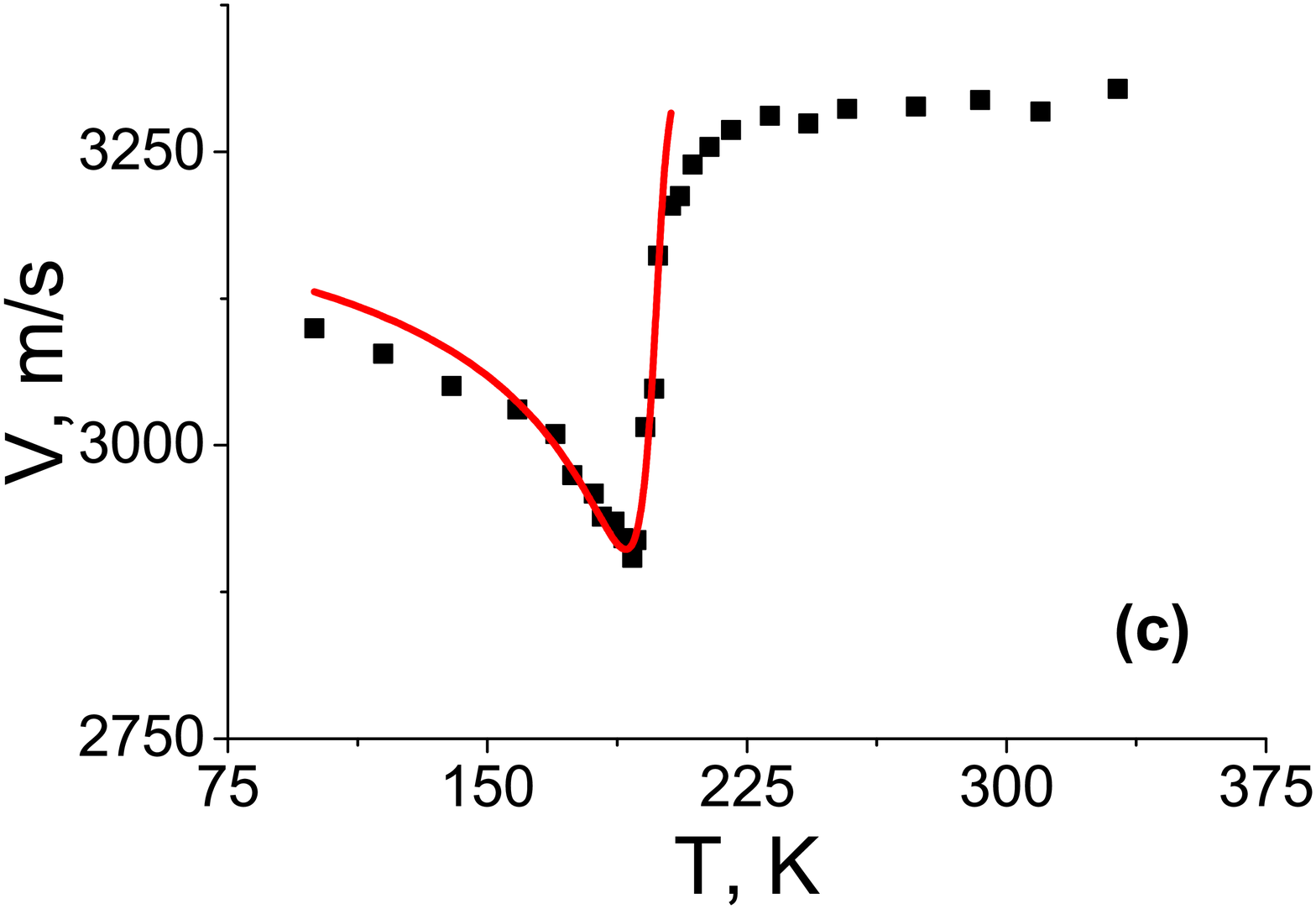}%
\includegraphics*[scale=0.23]{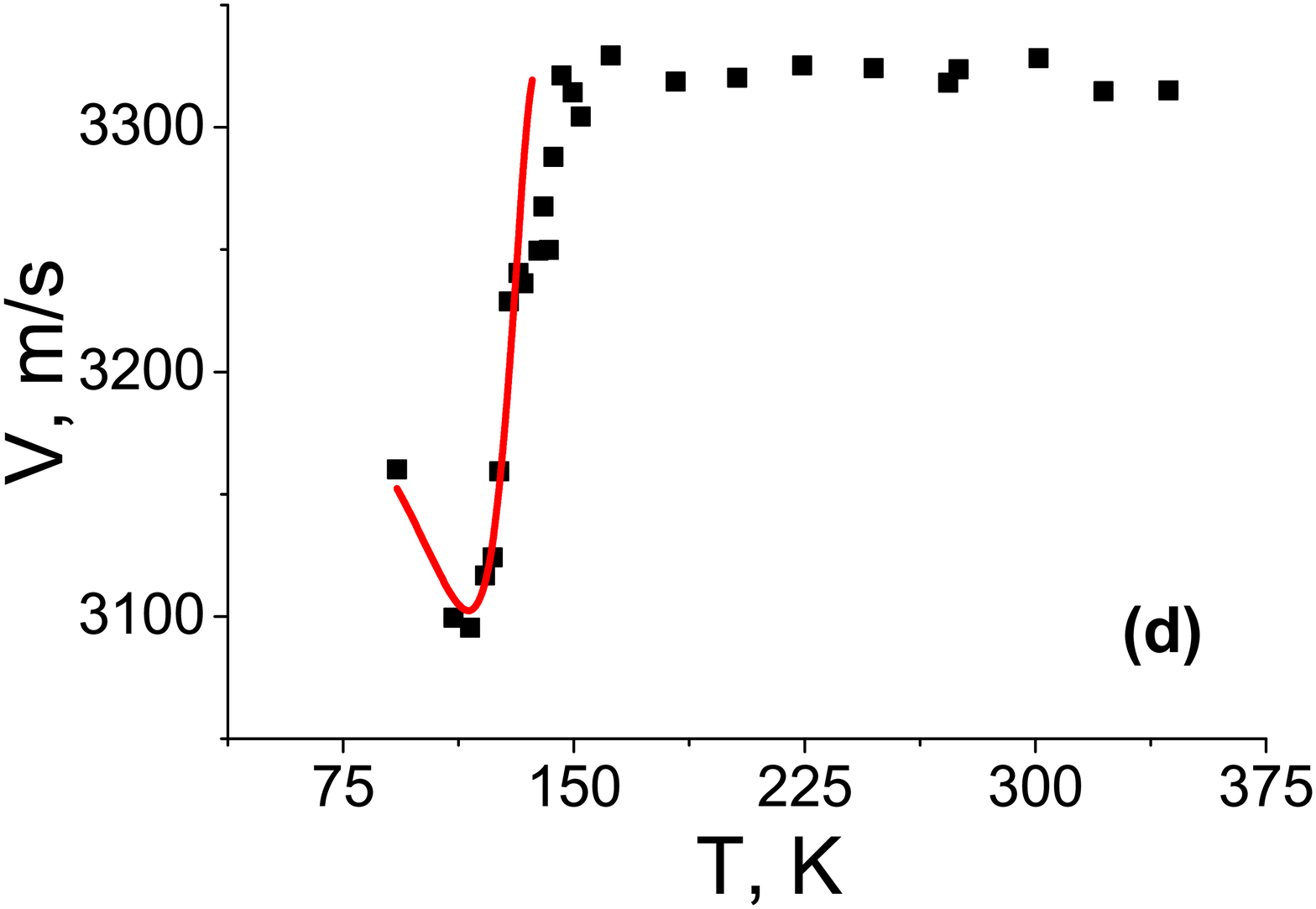}
\caption{Temperature dependencies of hypersound velocity for \PSPS\,  crystals (symbols) and fitting of their anomalies (lines) by Landau-Khalatnikov model (\ref{eq_v}): (a) -- $y=0$; (b) -- $y=0.2$; (c) -- $y=0.3$; (d) -- $y=0.45$.} \label{fig8}
\end{figure}

Using the Landau theory of second order PT, one can describe the sound velocity and attenuation anomalies. For the case of one-component order parameter and its interaction with strain, the thermodynamical potential could be presented in the following form:
\begin{equation}
\Phi=\Phi_0+\frac12\alpha P^2+\frac14\beta P^4+\frac16\gamma P^6+c_{ij}u_iu_j+q_{11i}u_{i}P^2+r_{11ij}u_iu_jP^2+...\,.
\label{eq_term_poten}
\end{equation}
\noindent Here $\alpha=\alpha_T(T-T_0)$, $\beta$ and $\gamma$ don't depend on temperature, $c_{ij}$ are elastic moduli, $q_{ijk}$ are electrostriction coefficients, $r_{ijkl}$ are bequadratic electrostriction coefficients. In approximation of one relaxation time $\tau=\tau_0/(T-T_0)$ for the order parameter dynamics, what was supposed in the Landau-Khalatnikov model~\cite{b20}, the following expressions for the temperature dependence of sound velocity and attenuation could be obtained:
\begin{equation}
V^2_{ij}=V^2_{ij\infty}-\frac{1}{1+\omega^2\tau^2}\Biggl[\frac{2q_{11i}q_{11j}}{\rho\beta\sqrt{1-\frac{4\alpha\gamma}{\beta^2}}}+
\frac{r_{11ij}\beta}{2\gamma\rho}\left(\sqrt{1-\frac{4\alpha\gamma}{\beta^2}}-1\right)\Biggr]\,, \label{eq_v}
\end{equation}
\begin{equation}
\alpha=\frac{V_{\infty}^2-V^2}{2V^3}\omega^2\tau\,.\label{eq_alpha}
\end{equation}
\begin{figure}[!htbp]
\centering
\includegraphics*[scale=0.23]{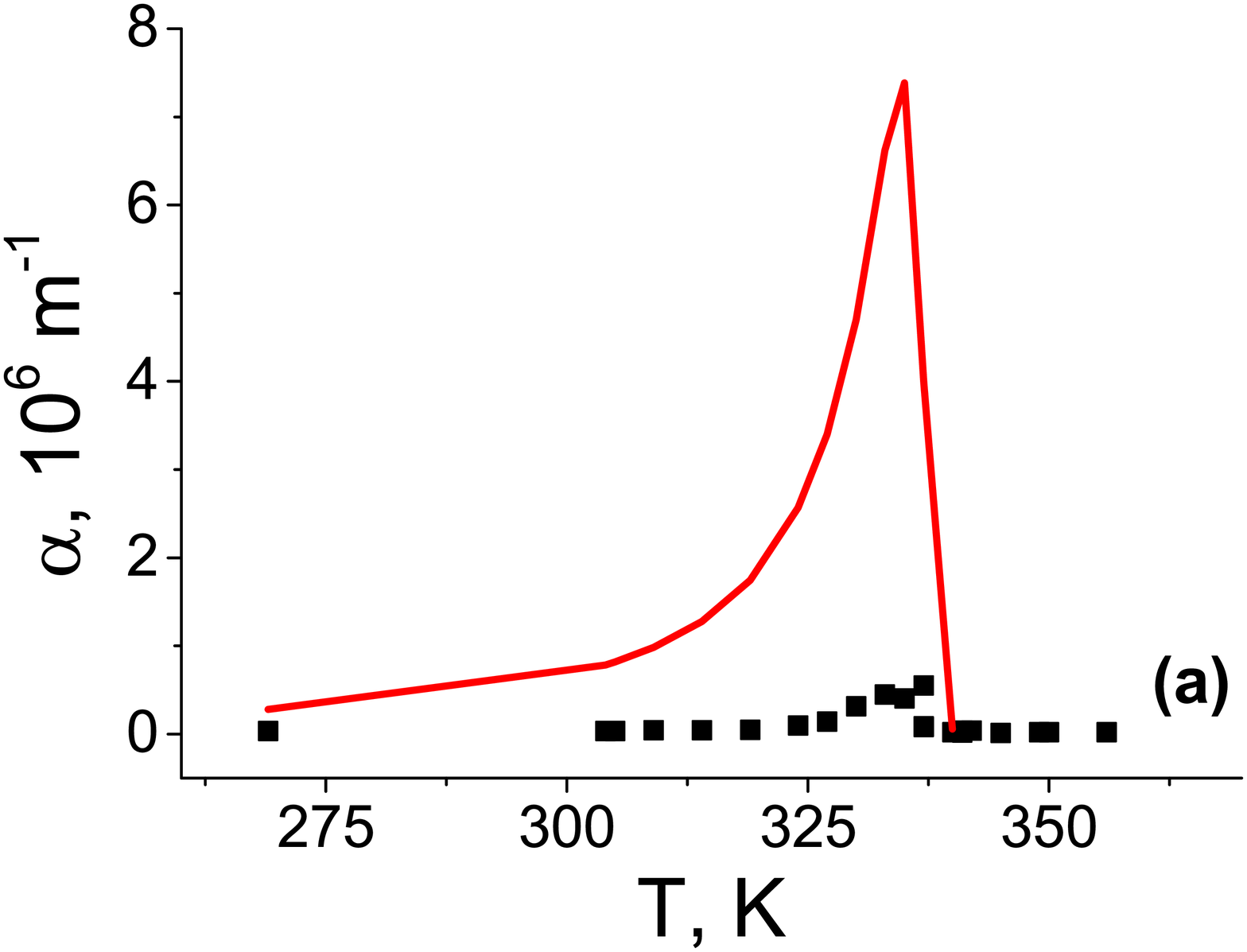}%
\includegraphics*[scale=0.23]{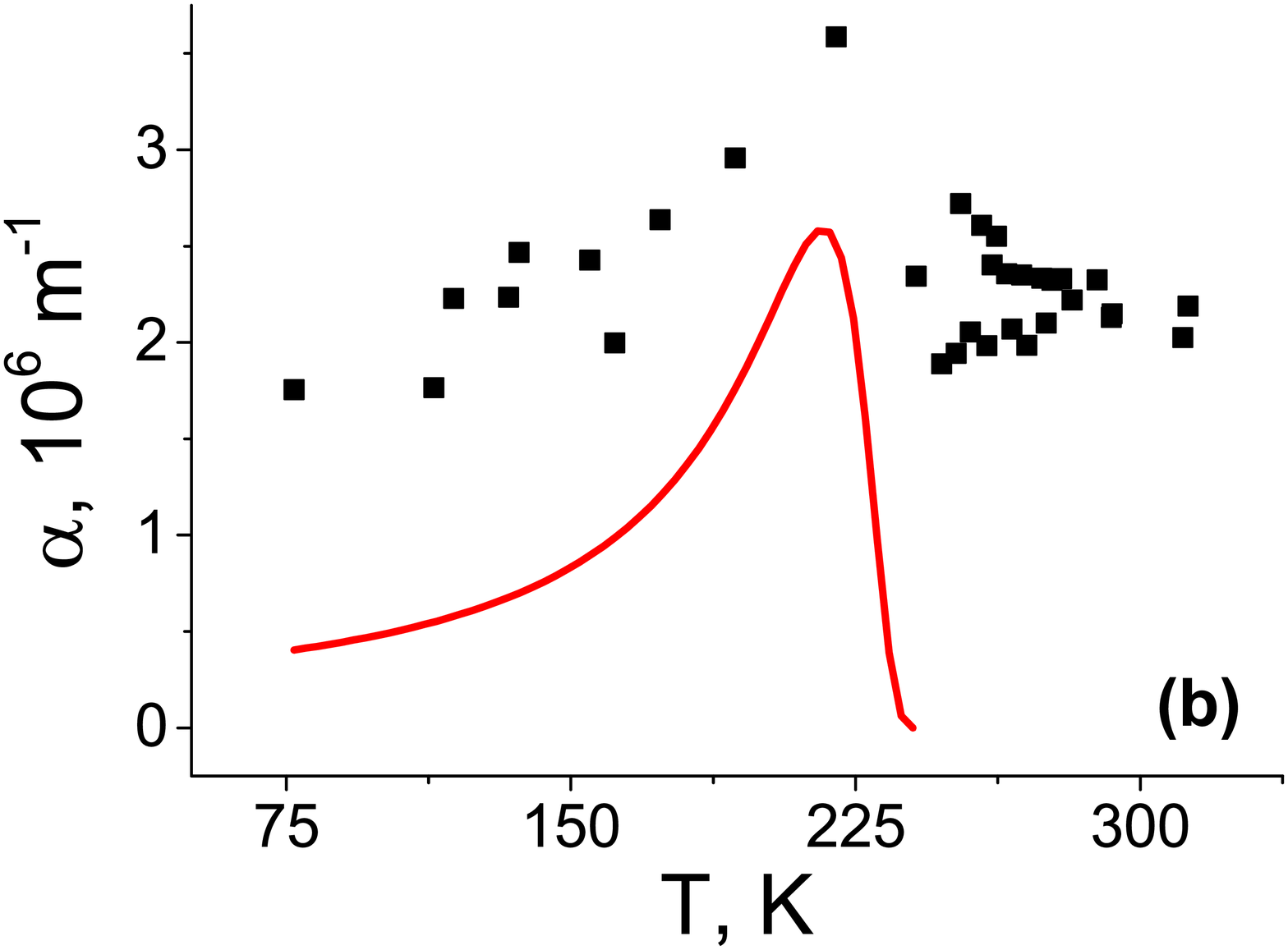}
\includegraphics*[scale=0.23]{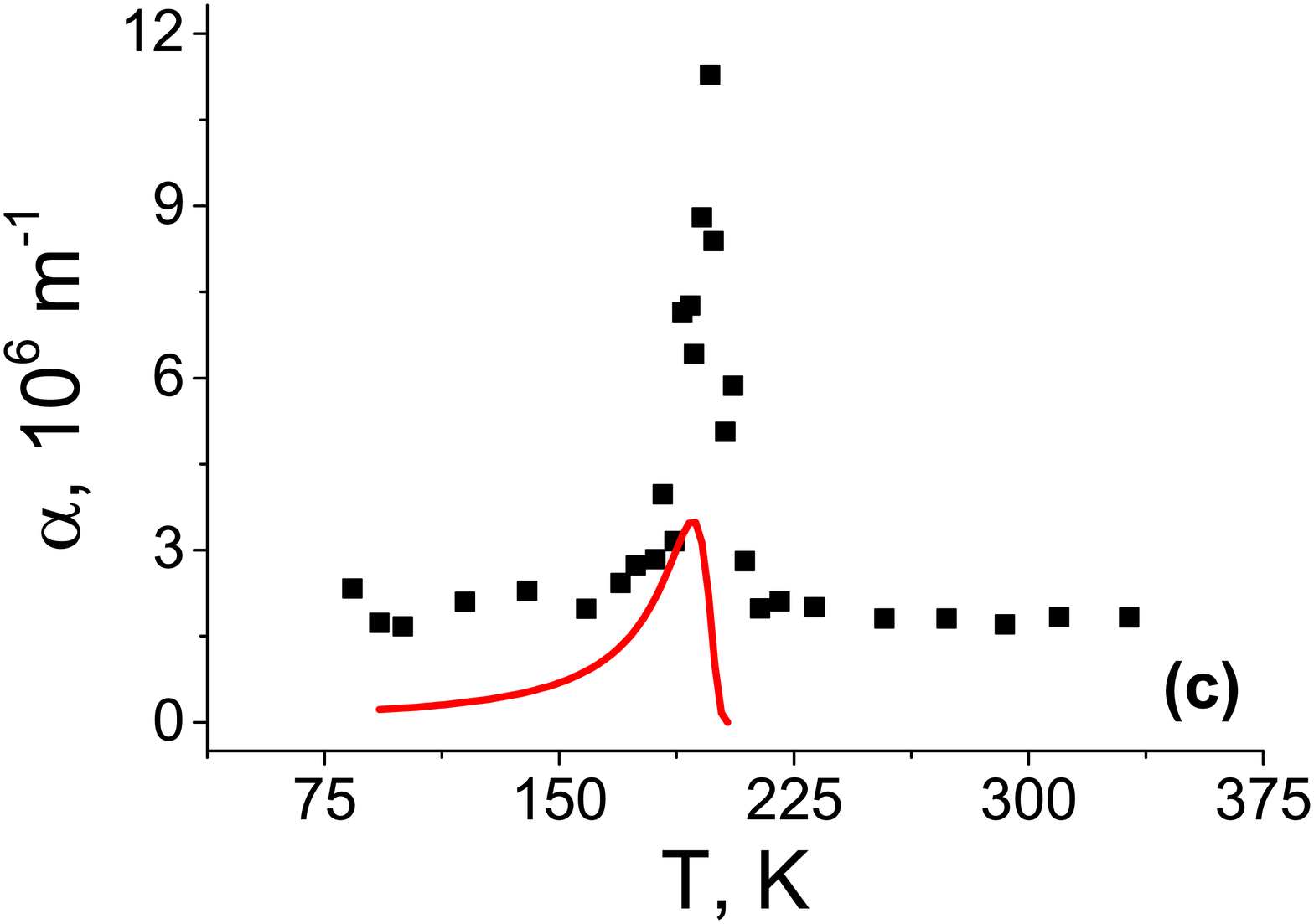}%
\includegraphics*[scale=0.23]{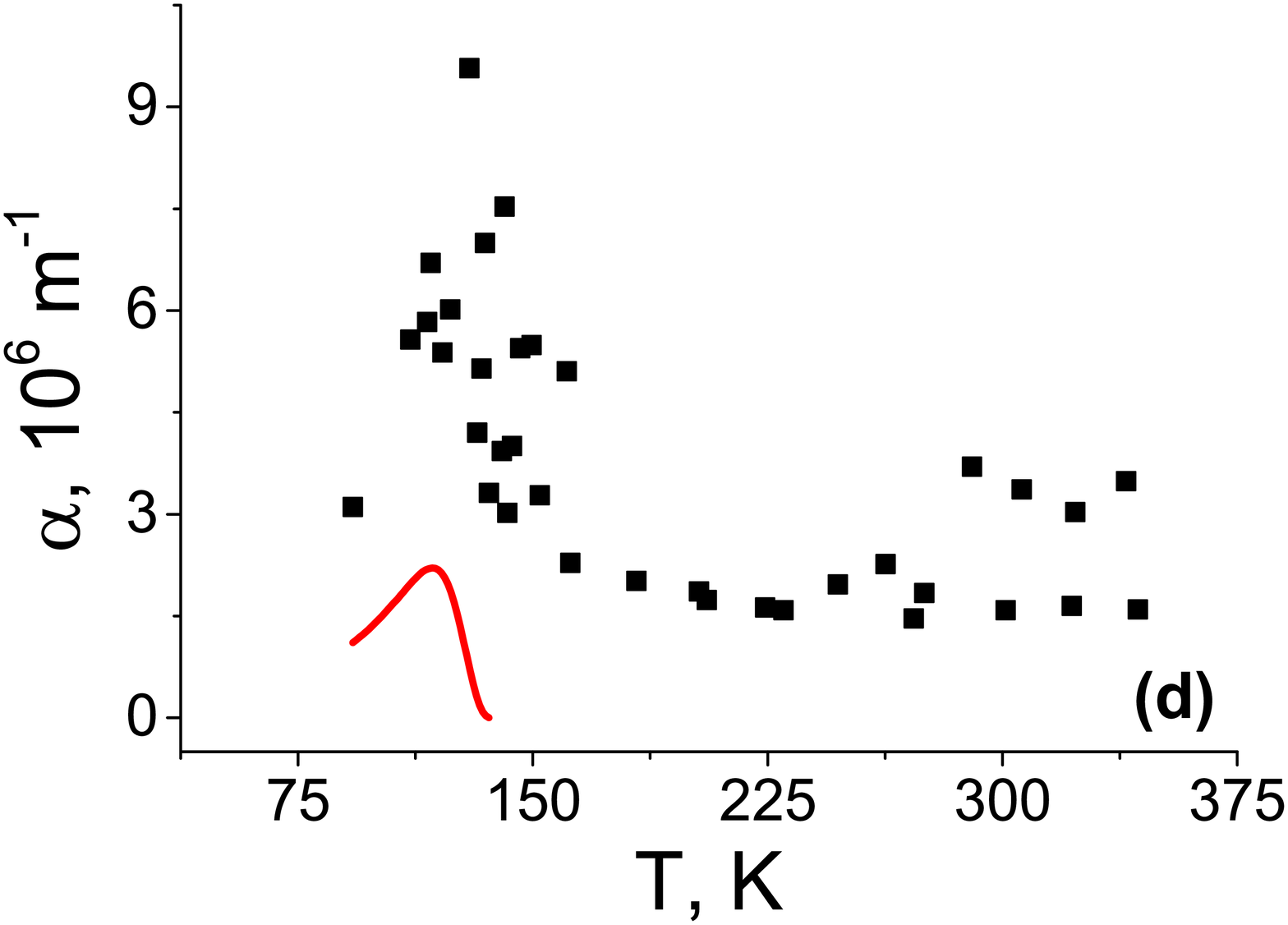}
\caption{Temperature dependencies of hypersound attenuation for \PSPS crystals (symbols) and their anomalies fitting (lines) by Landau-Khalatnikov model (\ref{eq_alpha}): (a) -- $y=0$; (b) -- $y=0.2$; (c) -- $y=0.3$; (d) -- $y=0.45$.} \label{fig9}
\end{figure}
\begin{figure}[!htbp]
\centering
\includegraphics*[scale=0.23]{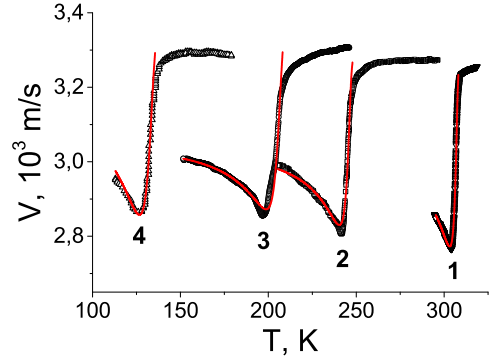}
\caption{Temperature dependencies of ultrasound velocity for \PSPS\, crystals (black symbols) and their anomalies fitting (red lines) by Landau-Khalatnikov model (\ref{eq_v}): 1-- $y=0.1$; 2 -- $y=0.2$; 3 -- $y=0.3$; 4 -- $y=0.45$.} \label{fig10}
\end{figure}

The analysis of temperature dependence of the longitudinal sound velocity propagating along the [010] direction in the \PSPS crystals (\fref{fig8} and \fref{fig10}) using of equation~(\ref{eq_v}) have been performed with the values of the coefficients $\beta$ and $\gamma$ for thermodynamic potential~(\ref{eq_term_poten}) shown on \fref{fig12}. It should be noted that $\alpha_T$ coefficient is almost independent on concentration and its value is about $1\times10^6$JmC$^{-2}$K$^{-1}$ and $1.25\times10^6$JmC$^{-2}$K$^{-1}$ from hypersound and ultrasound experiments respectively. It was found that for all compositions at description of hypersound velocity anomalies the electrostriction characteristics $q_{112}=3.4\times10^9$~J${\cdot}$m${\cdot}$C$^{-2}$ and $r_{1122}=0.1\times10^{10}$~N${\cdot}$m$^2$${\cdot}$C$^{-2}$ could be used. For fitting of ultrasound velocity anomalies the values $q_{112}=3.2\times10^9$~J${\cdot}$m${\cdot}$C$^{-2}$ and $r_{1122}=6.6\times10^{10}$~N${\cdot}$m$^2$${\cdot}$C$^{-2}$  have been used. The relaxation time temperature dependence is characterized by the values of $\tau_0$ coefficient that are also shown on~\fref{fig12}.

Calculated by relation~(\ref{eq_alpha}) ultrasound and hypersound attenuation in the ferroelectric phase (\fref{fig9} and \fref{fig11}) for all compositions of mixed crystals is higher than experimentally observed attenuation. This overestimation could be corrected by accounting of several relaxation times and mode Gruneisen coefficients. These coefficients are more appropriate as interacting parameters instead of electrostrictive coefficients as was earlier discussed in details for pure \SPS crystals~\cite{b21}.
\begin{figure}[!htbp]
\centering
\includegraphics*[scale=0.23]{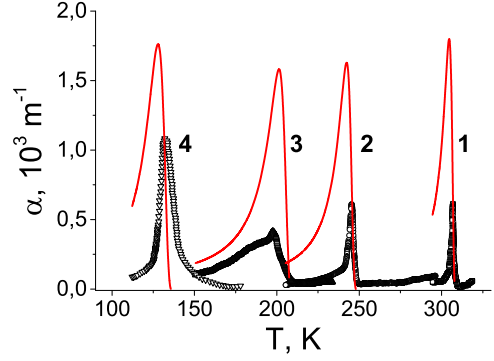}
\caption{Temperature dependencies of ultrasound attenuation for \PSPS\, crystals (black symbols) and their anomalies fitting (red lines) by Landau-Khalatnikov model (\ref{eq_alpha}): 1 -- $y=0.1$; 2 -- $y=0.2$; 3 -- $y=0.3$; 4 -- $y=0.45$.} \label{fig11}
\end{figure}
\begin{figure}[!htbp]
\centering
\includegraphics*[scale=0.23]{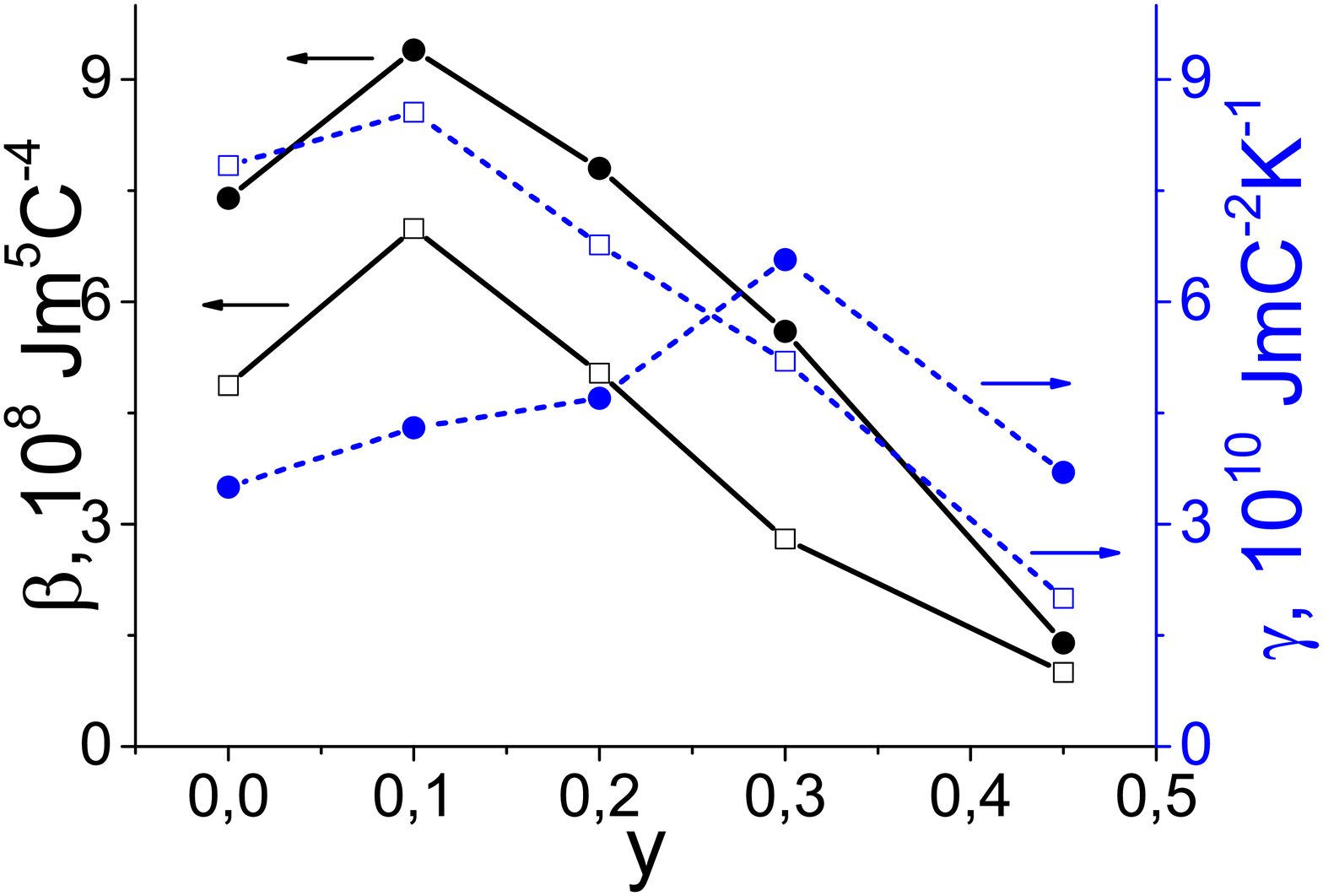}%
\hspace{0.1cm}\includegraphics*[scale=0.23]{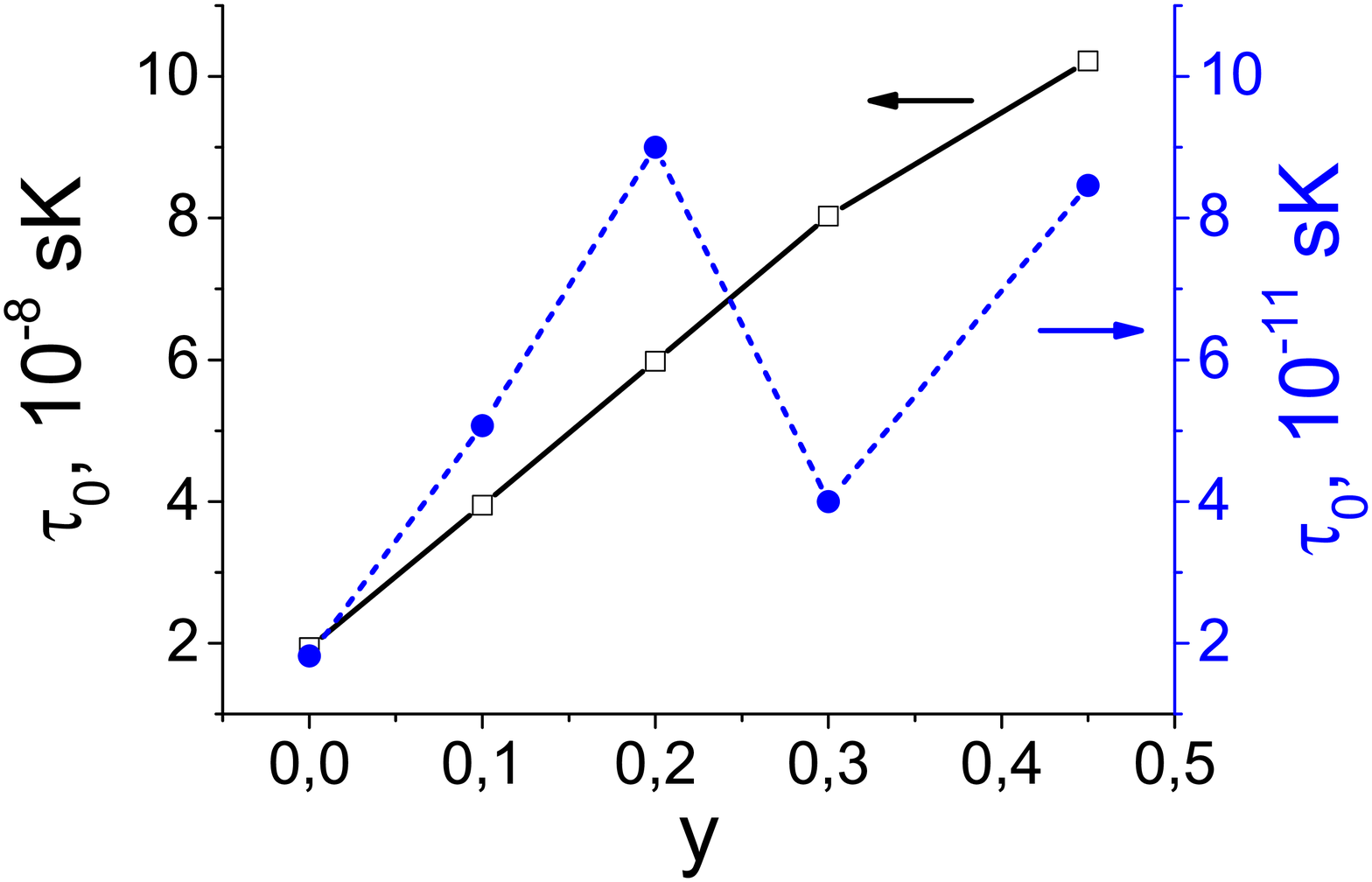}
\caption{Concentration dependencies of $\beta$ and $\gamma$ coefficients of thermodynamic potential (\ref{eq_term_poten}) and prefactor $\tau_0$ in the relaxation time temperature dependence founded at analysis of ultrasound  (open symbols) and hypersound (solid symbols) velocities and attenuation temperature anomalies in the Landau-Khalatnikov model (\ref{eq_v},\ref{eq_alpha}). } \label{fig12}
\end{figure}

\section{Discussion of results}
Anomalous growth of hypersound attenuation at $10^9$~Hz for $y=0.3$ composition (\fref{fig9}) could be related to enough small and fast heterophase fluctuations in the vicinity of first order ferroelectric phase transition. For $y=0.45$ sample, the heterophase fluctuations become more spaced and slower and they are involved into relaxation processes at lower frequencies what is demonstrated by strong growth of the dielectric losses at $10^4$~Hz (\fref{fig7}) and ultrasound attenuation at $10^7$~Hz (\fref{fig11}).

In addition, the relaxation time increasing at tin by lead substitution (\fref{fig12}) is obviously connected with strongly anharmonic order parameter dynamics in local three-well potential for phase transitions near the TCP. This is reached between $y=0.2$ and 0.3 and at temperature about 240~K.

\begin{figure}[!htbp]
\centering
\includegraphics*[scale=0.35]{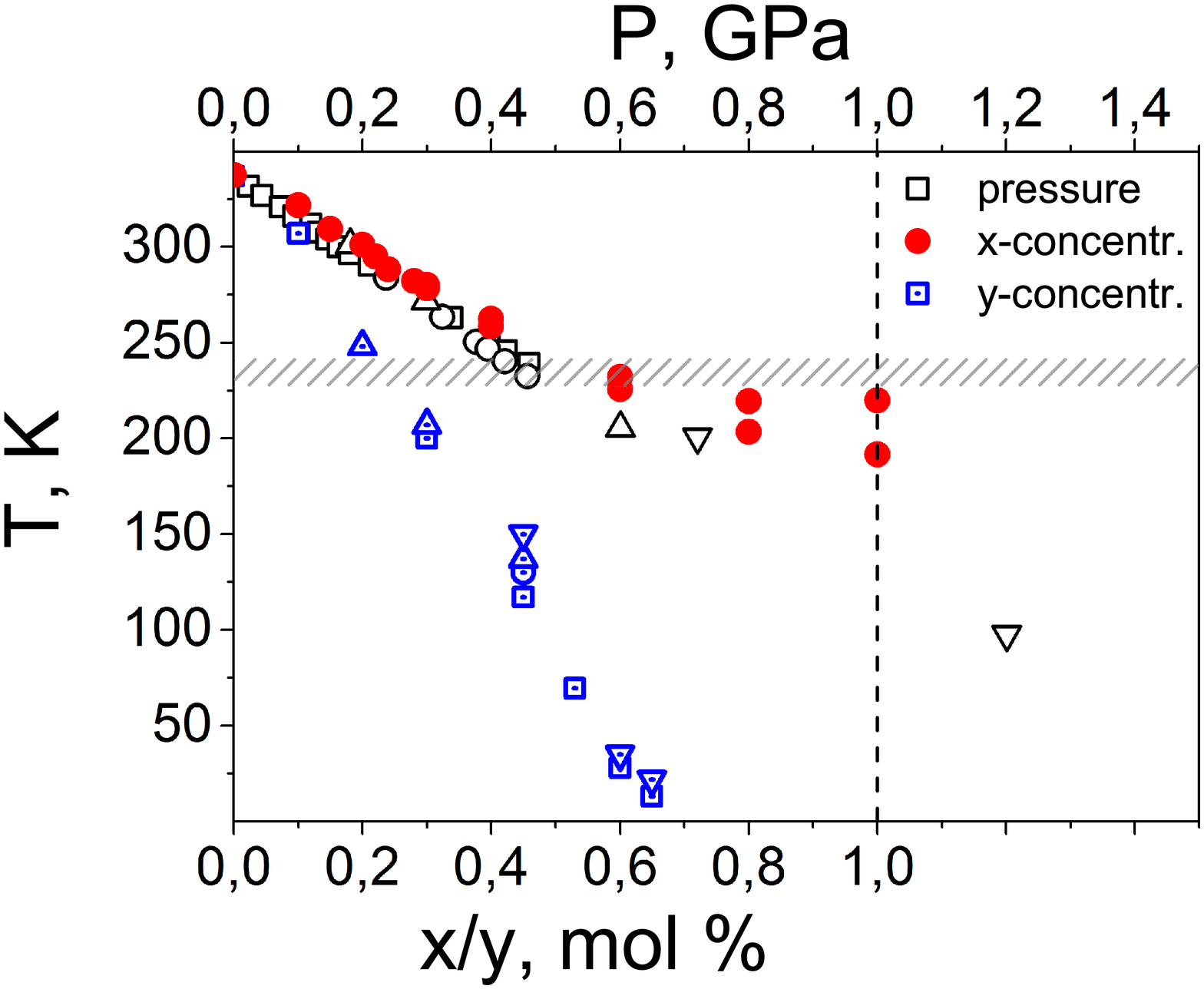}%
\caption{Open symbols -- pressure dependence of phase transition in \SPS\, crystals on the neutron diffraction~\cite{b8} and X-ray diffraction~\cite{b9} data. Solid symbols -- concentration dependence of phase transitions in \SPSSe\, crystals with intermediate incommensurate phase at $x>0.28$~\cite{b10,b11}. Doted symbols -- concentration dependence of phase transitions temperature in \PSPS\, crystals. Temperature positions of two dielectric and ultrasound anomalies are shown for compositions $y=0.3$ and $y=0.45$. For $y=0.45$ composition also the range of temperature hysteresis and possible phase coexistence are shown. For compositions with $y=0.53$, 0.61 and 0.66 the position of dielectric anomalies are shown according data~\cite{b15}. Horisontal shaded line shows "temperature waterline" that coincides with tricritical points positions on T--P, T--x and T--y diagrams. The vertical dashed line marks maximal values of compositions $x=y=1$.} \label{fig13}
\end{figure}

The TCPs location on T--P diagram for \SPS compound, on T--x diagram for \SPSSe, and T--y one for \PSPS mixed crystals could be compared (\fref{fig13}). For \SPS crystal the TCP is obviously located at $P\approx0.4$~GPa and $T\approx250$~K~\cite{b7,b8,b9}, for \SPSSe solutions the virtual TCP (inside of incommensurate phase) was predicted to be placed at $x\approx0.6$ and $T\approx240$~K~\cite{b10,b11}. It is intriguing that for \PSPS mixed crystals the TCP also could be placed at similar temperature level -- near 240~K at $y>0.2$.

For diluted BEG model, some segment of former first order transitions line (just below TCP in ideal system) becomes induced continuous transitions line by randomness~\cite{b5}. However, at further deviation of the phase transition point to lower temperatures, clear first order transition nature appears. The traditional mean field analysis of the thermodynamic potential coefficients concentration dependence, as was performed for \SPSSe mixed crystals~\cite{b10,b11}, could be complicated for \PSPS solutions. Indeed, in the latter case the concentration dependence of the coefficients is nonlinear.

As has been shown earlier~\cite{b22,b23} by analysis of \SPSe crystals heat capacity, dielectric susceptibility and wave number of modulation temperature dependence across the incommensurate phase, the higher order invariants (eight and ten) in the Landau expansion should be considered. Such nonlinearity is related to the presence of the three-well local potential. Obviously, these higher order invariants or linear temperature dependence of coefficient $\gamma$~\cite{b23} could be accounted for better fitting of the sound velocity temperature anomalies in \PSPS crystals and to avoid of the $\beta$ and $\gamma$ coefficients nonmonotonic dependence on the lead concentration (\fref{fig12}).

Observed location of TCPs at same temperature level ("temperature waterline") for different mechanical or chemical impacts (at compression of \SPS and at S by Se or Sn by Pb substitution) (\fref{fig13}) could be explained by the following way. At sulfur by selenium replacement, the intercell interactions become weaker in result of more covalent chemical bonds~\cite{b24,b25}, but the Sn$^{2+}$ cations stereoactivity and shape of the local potential obviously remain almost unchanged. These factors, mostly the intercell interactions weakening, determine the second order PT temperature lowering until reaching the TCP.

For pure \SPS crystal under pressure, the Sn$^{2+}$ cations stereoactivity decreases that lowers deepness of side wells in the local three-well potential. By this factor at almost unchanged intercell interactions, the second order PT temperature goes down until TCP "temperature waterline" and on T--P diagram the line of first order ferroelectric transition appears.

At tin by lead substitution, the sublattice of enough strongly stereoactive Sn$^{2+}$ cations is diluted by weakly stereoactive Pb$^{2+}$ cations, which have also bigger ionic radius. In addition to the dilution effect, the ionicity of Sn--S chemical bonds rise~\cite{b24} and stereoactivity of remained Sn$^{2+}$ cations lowers. Combined influence of the intercell interactions strengthening and two cations sublattices stereoactivity lowering determines decrease of the second order ferroelectric PT temperature down to already named "temperature waterline" of TCPs.

For pure \SPS lattice compression, the observed T--P phase diagram could be related to the predicted one in the BEG model~\cite{b2,b3,b4} and to founded by MC simulations (\fref{fig1}) on the basis of first-principles evaluated local three-well potential for this ferroelectrics~\cite{b1}. At sulfur by selenium substitution, the random field defects could appear but their influence is smoothed -- very sharp critical anomalies at the phase transitions were observed in the \SPSSe mixed crystals~\cite{b26}. This is obviously related to continuous electronic density space distribution in [P$_2$S(Se)$_6$]$^{4-}$ anion groups. For these mixed crystals obviously, the BEG model could be appropriate in combination with known ANNNI model~\cite{b27,b28} that explains frustration effects and intermediate incommensurate phase appearance at $x>x_{LP}\approx0.28$~\cite{b14}.

For the case of tin by lead replacement, a strong random field defects appear on the matter of different electronic orbitals hybridization around Sn$^{2+}$ and Pb$^{2+}$ cations. Such complicate situation could be described at compare of experimentally founded T--y diagram with predicted diagram for diluted BEG model~\cite{b5}.

\section{Conclusions}
The temperature-pressure phase diagram has been investigated by MC simulations for \SPS ferroelectrics with local three-well potential. At compression, the temperature of second order transition from paraelectric phase into ferroelectric one lowers to TCP near 250~K and 0.4~GPa. At further compression and for lower temperatures, the quadrupolar and disordered phases could appear. The TCP below 250~K and between 0.4 and 0.6~GPa experimentally was observed in \SPS crystals as appearance of clear first order transition~\cite{b8,b9}. At similar temperature level, or near "temperature waterline" below 250~K, the TCP is appeared for paraelectric-ferroelectric transition in \SPSSe mixed crystal with selenium content $x\approx0.6$~\cite{b10,b11}. By hypersound, ultrasound and dielectric investigations of \PSPS mixed crystals, it have been found that TCP could be also reached at the phase transition temperature lowering to the "temperature waterline" below 250~K while lead concentration increase more than $y=0.2$. At higher content of lead, the wide temperature hysteresis of the phase transitions and, obviously, phases coexistence are observed. The \PSPS mixed crystals represent disordered ferroelectric system that could be described by BEG model with random fields.

\end{document}